\documentclass{article}
\usepackage[no-natbib]{jheppub}
\usepackage{graphicx} 
\usepackage{xcolor} 
\usepackage{tabularx}
\usepackage{amsmath,amssymb,amsthm,amsfonts,bbold}
\usepackage{physics}
\usepackage{subcaption}
\usepackage[sorting =none,
          maxnames=1000,
          backend=biber
          ]{biblatex}
\hypersetup{linkcolor=red}

\newcommand{\Hg}{\mathcal H_{\gamma}}
\newcommand{\Ha}{\mathcal H_{a}}

\newcommand{\SL}{S_{\mathrm L}}

\title{\boldmath New quantum information perspectives in the axion--photon and neutrino systems}

\author[a,b]{Aaditya Datar,}
\author[a]{Arun M. Thalapillil}
\author[c]{and Palak Thareja}
\affiliation[a]{Indian Institute of Science Education and Research, Pune\\Dr. Homi Bhabha Road,
Pune 411008, India}
\affiliation[b]{Quantum and Advanced Technologies Research Institute\\Griffith University, Yuggera Country, Brisbane, QLD 4111, Australia}
\affiliation[c]{Miranda House\\University of Delhi, North Campus, Delhi 110007, India}

\emailAdd{aaditya.datar@griffithuni.edu.au}
\emailAdd{thalapillil@iiserpune.ac.in}
\emailAdd{palak.2023.876@mirandahouse.ac.in}

\abstract{
In this work, we broach a quantum information-theoretic treatment of axion--photon mixing. Motivated by an emerging class of quantum-enhanced axion searches, we analyse the two-level single-excitation sector of axion--photon oscillations, demonstrating how the coupled dynamics naturally generate bipartite axion--photon mode entanglement. We study in detail the ensuing aspects of entanglement entropy, concurrence, negativity, quantum mutual information and discord, and capacity of entanglement, and the corresponding neutrino analogues wherever novel and previously unaddressed. In particular, we highlight the characteristic features that connect maximal axion--photon entanglement to resonant or strong-mixing conversion, and the distinct thresholds for the extremal values attained by the quantum information measures. We study aspects of the Mandelstam--Tamm and Margolus--Levitin quantum speed limits for both the axion--photon and neutrino systems. While orthogonalisation occurs only at axion--photon resonance, or at maximal neutrino mixing, where the two bounds coincide, away from these limits, the Margolus--Levitin bound is saturated at maximal conversion, while the Mandelstam--Tamm bound is generally weaker. We also study an entanglement quantum speed limit for axion--photon conversion, that separates into detuning-dominated and magnetic-mixing-dominated regimes, and find that it is saturated for a period and then the bound becomes weak. The results in this work identify the quantum resources and limiting timescales intrinsic to axion--photon conversion, and connect axion phenomenology, neutrino oscillations and quantum information theory.
}
 
\begin{document}

\maketitle
\section{Introduction}\label{sec:introduction}

Axions and axion-like particles remain among the most compelling extensions of the Standard Model, motivated by the strong-$CP$ problem and by the possibility that light, weakly coupled pseudoscalars constitute part of the dark sector\,\cite{Peccei:1977hh, Weinberg:1977ma, Wilczek:1977pj}. Their coupling to photons, in an external electromagnetic background, leads to coherent axion--photon interconversion\,\cite{Primakoff:1951iae}. This phenomenon admits a two-state mixing description\,\cite{PhysRevD.37.1237} and underlies a broad class of cavity, helioscope, haloscope and interferometric searches\,\cite{CAST:2004gzq,CAST:2007jps,Irastorza:2018dyq,DiLuzio:2020wdo,Caputo:2024oqc}. The increasing role of quantum measurements, quantum sensing and single-photon techniques in such searches\,\cite{Degen:2017QuantumSensing,Dixit:2020ymh,Yang:2022uil,Yu:2023wsn,Braggio:2024xed,Zheng:2024kxn,Devlin:2026czx} prompts a complementary description of the axion--photon system in the language of quantum information. This is among the motivations for the present work.

In the context of this study, the relevant quantum regime is naturally described in the single-excitation sector. Here, the Hilbert space is spanned by a photon excitation, in the polarisation mode parallel to the external magnetic field, and an axion excitation, so that coherent oscillations delocalise one quanta across two field modes. This is an example of mode entanglement and is familiar from discussions of single-particle entanglement in the context of quantum optics\,\cite{Tan:1991oqb,Yurke:1992qpa,PhysRevA.64.042106,PhysRevA.72.064306,Zanardi:2001zz,Zanardi:2002pwt,PhysRevLett.92.180401,Fuwa:2014cra,Guerreiro:2016ytu}. The axion--photon entanglement is analogous in form to entanglement analyses of two-flavor neutrino oscillations\,\cite{Blasone:2007wp,Blasone:2007vw}, but differs in physical content---in contrast to neutrino oscillations, the axion--photon oscillations persist classically, and their dependence on physical parameters is conceptually distinct.

In this paper, we develop a quantum information-theoretic analysis of axion--photon oscillations, with comparisons to the two-flavour neutrino system wherever appropriate. We investigate a comprehensive list of quantum information measures in both the interaction basis and in the propagation basis, and detail their characteristic behaviour in the axion--photon system. Although the structure of the single-excitation sector implies that the standard entanglement, discord, and coherence measures are all functions of a single Schmidt weight\,\cite{Vidal:1998re}, and hence of the axion--photon conversion probability, the measures still provide complementary perspectives. For fixed physical parameters, we point out that the extrema attained by the various quantum information measures have very distinct thresholds. 

We note various characteristic features of the quantum information measures---for instance, the maxima of various standard quantum information measures may occur at the axion--photon resonance condition, where the effective photon mass equals the axion mass, or are approached in the regime of dominant magnetic mixing. These scenarios are  effectively within the regime of enhanced conversion that are targeted by present axion search strategies\,\cite{CAST:2007jps,Irastorza:2018dyq,DiLuzio:2020wdo,Caputo:2024oqc} and it will be interesting to explore how quantum information perspectives discussed in the present work translate realistically to novel experimental methods\,\cite{Degen:2017QuantumSensing,Dixit:2020ymh,Yang:2022uil,Yu:2023wsn,Braggio:2024xed,Zheng:2024kxn,Devlin:2026czx}. Overall, since the dynamical measures are mainly functions of the conversion probability, they may in principle be inferable from single-excitation signal statistics in quantum-enhanced axion searches\,\cite{Dixit:2020ymh,Yu:2023wsn,Braggio:2024xed,Devlin:2026czx}.

Quantum discord provides a diagnostic of non-classical correlations in bipartite systems, thereby extending beyond entanglement-based characterizations\,\cite{Ollivier:2001fdq, Brodutch:2016lvw, Modi:2012baj}. Operationally, it captures the genuinely quantum disturbance associated with local measurements, which has no analogue in classical correlated systems. In the axion--photon system, the equality between discord and entanglement entropy is a special feature of the pure two-mode states considered here and is not expected to persist once realistic sources of decoherence are included. With an eye towards more experimentally realistic studies in the future, leveraging quantum information-theoretic measures in axion--photon searches, we study discord.

The capacity of entanglement\,\cite{Yao:2010woi,Schliemann:2011zkg} gives an important exception to the degeneracy among the re-parametrised monotones---it probes fluctuations of the entanglement spectrum rather than its magnitude, and thus, is conceptually different from the other entanglement measures. In the axion--photon system it exhibits a characteristic twin-peaked profile, with its global maximum at a non-trivial conversion probability rather than at maximal entanglement; a corresponding observation is also made in the neutrino case. Later, we also investigate the role of the capacity of entanglement in setting a quantum speed limit for the generation of entanglement entropy\,\cite{Robertson1929,Shrimali:2022bvt,Thakuria:2022taf,Mohan:2021rel}.

We further analyse axion--photon and neutrino oscillations from the perspective of quantum speed limits (QSLs). For orthogonalisation, the Mandelstam--Tamm and Margolus--Levitin bounds\,\cite{Mandelstam1991,Margolus:1997ih,LevitinToffoli2009,Deffner:2017cxz} are attainable only in restricted regimes---maximal vacuum mixing for the two-flavour neutrinos and the axion--photon resonance condition for the single-excitation axion--photon system. In these cases, the two bounds coincide, as expected from a general theorem for equal-weight two-level superpositions of energy eigenstates\,\cite{LevitinToffoli2009}. It is also noted that the parameter regimes coincide with maximal mode entanglement. An intriguing aspect also pertains to the role of $\hbar$ in the two systems---while the neutrino orthogonalisation quantum speed limit vanishes in the classical limit, reflecting the intrinsically quantum-mechanical character of neutrino flavour oscillations, the resonant axion--photon characteristic time is independent of $\hbar$, and survives as a scale characterising the classical coupled-wave conversion. In the classical regime, it therefore admits the interpretation of a minimum spatial scale for interconversion between two classical field modes, rather than a purely quantum-mechanical timescale. 

Away from exact orthogonalisation, we study the generalised Mandelstam--Tamm and Margolus--Levitin bounds for non-orthogonal evolution. At the oscillation half-period, the generalised Mandelstam--Tamm time is finite but not saturated in either system, whereas the generalised Margolus--Levitin time is exactly saturated for both the neutrino and axion--photon oscillations throughout the parameter space. Consequently, in the axion--photon and neutrino systems, the combined Levitin--Toffoli quantum speed limit bound\,\cite{LevitinToffoli2009} is set by the Margolus--Levitin contribution\,\cite{Margolus:1997ih}.

Finally, we investigate a quantum speed limit for the generation of entanglement entropy\,\cite{Robertson1929,Shrimali:2022bvt,Thakuria:2022taf,Mohan:2021rel}. The bound is governed by the Hamiltonian variance and by the time-averaged square root of the capacity of entanglement. It is tight over an initial part of each oscillation period and then weakens. In the magnetic-mixing-dominated regime, the saturation interval is long and primarily controlled by the mixing strength---the bound has a characteristic double-humped profile. In the detuning-dominated regime, the saturation interval is shorter, only weakly sensitive to the mixing strength, and the profile is single-humped. These two behaviours correspond to distinct phenomenological regions of the axion search parameter space.

The paper is organised as follows. In Sec.~\ref{sec:axionphoton}, we review axion--photon mixing in an external magnetic field, relate the classical and single-excitation quantum regimes, and compare the resulting two-level structure with two-flavour neutrino oscillations. In Sec.~\ref{sec:axionphotonQuantumMeasures}, we formulate the mode entanglement structure and investigate in detail the various quantum information measures of interest. In Sec.~\ref{sec:axionphotonQSLs}, we comprehensively study the Mandelstam--Tamm, Margolus--Levitin and entanglement quantum speed limits. We summarise and conclude in Sec.~\ref{sec:summaryconclusions}.

\section{Axion--photon oscillations}\label{sec:axionphoton}

\subsection*{Axion--photon mixing}
Consider the behaviour of the coupled axion--photon system in the presence of a classical background, specifically, a temporally and spatially constant external magnetic field $\vec{B}_{\text{ext}}$ transverse to the propagation direction (here, the $z$ direction). The external magnetic field is assumed to be much larger in magnitude than the dynamical magnetic field $\vec{B}_{\gamma}$ associated with the propagating photon ($|\vec{B}_{\text{ext}}| \gg |\vec{B}_\gamma|$). We will adopt natural units ($\hbar=c=1$) throughout, restoring $\hbar$ occasionally to clarify concepts as required. The relevant equations of motion for the scalar axion $a(z)$, and the components of the photon field transverse ($\vec{A}_{\perp}(z)$) and parallel ($\vec{A}_{\parallel}(z)$) to $\vec{B}_{\text{ext}}$ are\,\cite{PhysRevD.37.1237} 
\begin{equation}\label{eq:propeq}
    \left( \omega^2 + \partial_z^2 \right) \mathbb{\Psi}(z) = \mathbb{M}^2 \cdot \mathbb{\Psi}(z) \; ,
\end{equation}
with the definitions
\begin{equation}
    \mathbb{\Psi}(z) = \begin{pmatrix}
        A_\perp(z)\\
        A_{\parallel}(z)\\
        a(z)
    \end{pmatrix} \;,\qquad\qquad
    \mathbb{M}^2 = \begin{pmatrix}
        m_{\gamma,\perp}^2 & 0 & 0 \\
        0 & m_{\gamma,\parallel}^2 & g_{a\gamma\gamma}\omega B_{\text{ext}}\\
        0 & g_{a\gamma\gamma}\omega B_{\text{ext}} & m_a^2
    \end{pmatrix} \; .
\end{equation}
The mass parameters $m_{\gamma,\perp}^2$ and $m_{\gamma,\parallel}^2$ denote the possible modification of the photon dispersion relations due to the presence of the external magnetic field and/or the medium of propagation itself. Any Faraday rotation effects have been assumed to be subdominant. In this regime, one observes a mixing between the parallel component of the photon field and the axion field, while the transverse photon component decouples.

One may linearise Eq.\,\eqref{eq:propeq} and rewrite the relevant equation describing axion--photon oscillations as an effective two-state system, satisfying the mixing equation (see Appendix\,\ref{Appendix:AxionPhotonEquations} for details)

\begin{equation}\label{eq:schrodingereq}
    i \partial_z \Psi(z) = \mathbb{H}  \cdot \Psi(z) \; ,
\end{equation}
with
\begin{equation}\label{eq:hamiltonian}
    \Psi(z) = \begin{pmatrix} A_\parallel(z) \\ a(z) \end{pmatrix} \;,\qquad\qquad
    \mathbb{H} = \frac{1}{2\omega}
    \begin{pmatrix}
        m_{\gamma,\parallel}^2 & g_{a\gamma\gamma} \omega B_{\text{ext}} \\
        g_{a\gamma\gamma} \omega B_{\text{ext}} & m_a^2
    \end{pmatrix} \; .
\end{equation}
This matrix may be further decomposed into a mixing part and a term proportional to the identity
\begin{align}
    \mathbb{H} &= \frac{m_{\gamma,\parallel}^2 + m_a^2}{4\omega} \mathbb{1} +
    \begin{pmatrix}
        \alpha & \beta \\
        \beta & -\alpha
    \end{pmatrix} \; ,
\end{align}
where
\begin{equation}\label{eq:alphabeta}
    \alpha = \frac{m_{\gamma,\parallel}^2 - m_a^2}{4\omega}\;,\quad
    \beta = \frac{g_{a\gamma\gamma} B_{\text{ext}}}{2} \;.
\end{equation}
In our subsequent studies, we will express all the quantum information measures in terms of these constants ($\alpha$ and $\beta$) which naturally appear as combinations of the physical parameters.

The component proportional to the identity only contributes an overall common phase which does not affect the physical transition probabilities, and so, we focus on the part that contributes to the oscillations
\begin{equation}\label{eq:mixingmatrix}
    H = \begin{pmatrix} \alpha & \beta \\ \beta & -\alpha \end{pmatrix} \; .
\end{equation}
The oscillation probability may be calculated from the above to be\footnote{Note that this is calculated for an initial state that is purely photonic, converting to an axionic state. The expression remains identical for the complementary scenario due to the symmetry of the mixing. That is, the expression in Eq.\;\eqref{eq:osc_prob} is also the transition probability $P_{a \rightarrow \gamma}(z)$ for an initial state that is purely axionic, converting into a photonic state. A corollary is that the present study also applies to axion detection experiments utilising single-photon techniques\,\cite{Dixit:2020ymh,Yu:2023wsn, Braggio:2024xed,Devlin:2026czx}. }
\begin{equation}\label{eq:osc_prob}
P_{\gamma \rightarrow a}(z)=\frac{\beta^2}{\alpha^2+\beta^2}\sin^2(z\sqrt{\alpha^2+\beta^2}) \; .
\end{equation}
Note that the maximum conversion probability attained is $P^{\,\max}_{\gamma \rightarrow a}=\beta^2/(\alpha^2+\beta^2)$, and depends crucially on the physical parameters $\alpha$ and $\beta$. Consequently, the conversion probability does not span the full $[0,1]$ range in general.

\begin{figure}[h!]
    \centering
    \includegraphics[width=0.75\textwidth]{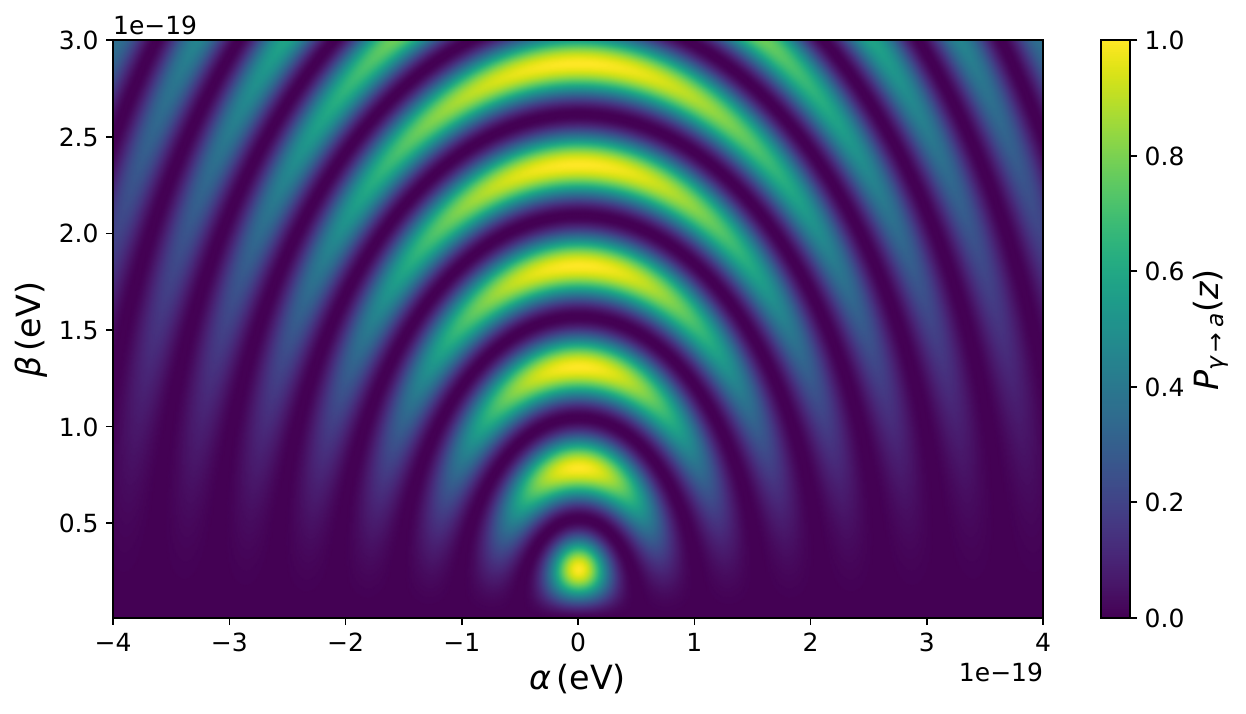}
    \caption{Representative plot of the transition probability $P_{\gamma\rightarrow a}(z)$, for a fixed $z =6 \times10^{19} \, \mathrm{eV}^{-1}$, merely to illustrate its dependence on the physical parameters $\alpha$ and $\beta$. Realistic values of the physical parameters $\alpha$ and $\beta$ lead to very small transition probabilities.}
    \label{fig:transitionprobability}
\end{figure}
Fig.\,\ref{fig:transitionprobability} illustrates the functional form of the transition probability as a function of $\alpha$ and $\beta$. The actual values of the expected transition probabilities for realistic values of the physical parameters $\alpha$ and $\beta$ are very small. For instance, for the baselength $\mathrm{L}\sim 10\,\mathrm{m}$ and magnetic field $B_{\text{ext}}\sim 10\,\mathrm{T}$ in the CAST experiment\,\cite{CAST:2004gzq}, and assuming $g_{a\gamma\gamma}\sim\,10^{-10}\,\mathrm{GeV}^{-1}$, one gets $P_{a \rightarrow \gamma}\sim \mathcal{O}(10^{-17})$ if one assumes that the oscillation length exceeds the telescopic length.

\subsection*{Classical and quantum regimes of the oscillations}
The coupled axion–photon system can be described in two distinct limits. At the level of classical field theory, as discussed before, the scalar axion and the photon field satisfy equations of motion that are linear classical wave equations in the presence of a classical background $\vec{B}_{\text{ext}}$, and a coupling between the fields characterised by $g_{a\gamma\gamma}$. In this regime, the system oscillations are described in terms of the classical field amplitudes.

We are primarily interested in a quantum information-theoretic treatment of the axion--photon system. Hence, our regime of interest is the one where a quantum mechanical description is more appropriate. Such a description becomes relevant when the occupation number of the photon and axion modes is small (and additionally, on an operational level, when measurement techniques are sensitive enough to distinguish between individual photons). In this low occupation number limit, the system is best represented by individual quanta of the fields and their coherent superpositions. The relevant Hilbert space of interest to us is therefore spanned by
\begin{equation}
|\gamma\rangle \equiv |1_\gamma,0_a\rangle \; , \qquad |a\rangle \equiv |0_\gamma,1_a\rangle \; .
\end{equation}
which correspond to a single photon or a single axion excitation, respectively\footnote{Note that although we will use the notation $|\gamma\rangle$, it is only associated with the $A_{\parallel}(z)$ excitation.}. We will further expand on this coherent superposition in the next section and motivate why this is an example of quantum mode entanglement, and therefore appropriate to be expressed in this occupation number notation that is suggestive of qubits.

The reduction to a two-level system in our particular scenario corresponds to restricting the full field theory to a single momentum mode, and to the single-particle sector as discussed in Appendix~\ref{Appendix:AxionPhotonEquations}. Quantitatively, we can approximately describe the quantum regime in terms of bosonic creation and annihilation operators for the photon and axion modes. Restricting to the photon polarisation that couples to the external magnetic field, an approximate effective Hamiltonian operator of the coupled system can then be taken as (restoring the $\hbar$ factors for clarity)
\begin{equation}\label{eq:effhamiltonian}
    \hat{H}= \frac{m^2_{\gamma,\parallel}}{2\hbar \omega}\;\hat{b}^{\dagger} \hat{b}
    + \frac{m^2_a}{2 \hbar \omega}\;\hat{c}^{\dagger} \hat{c}
    + \frac{\hbar g_{a\gamma\gamma}B_{\text{ext}}}{2}\left(\hat{b}^{\dagger} \hat{c}+\hat{c}^{\dagger} \hat{b}\right) \;,
\end{equation}
Here, $(\hat{b},\hat{b}^\dagger)$ and $(\hat{c},\hat{c}^\dagger)$ are the bosonic annihilation and creation operators for the photon and axion modes respectively, and the photon and axion states are created when $\hat{b}^\dagger, \hat{c}^\dagger$ act on $|0_\gamma,0_a\rangle$. One may provide a more thorough treatment of the quantum regime in the axion--photon system (see, for instance,\,\cite{Ikeda:2025qac} and Appendix~\ref{Appendix:AxionPhotonEquations}).

In this regime, a general state can be written as
\begin{equation}
|\psi(z)\rangle = c_1(z)\left|1_\gamma,0_a\right\rangle + c_2(z)\left|0_\gamma,1_a\right\rangle\equiv c_1(z)|\gamma\rangle + c_2(z)|a\rangle \; .
\end{equation}
The Schr\"odinger equation then describes the coherent superposition of a single quantum of excitation between the photon and axion modes as\footnote{We have gone from $z$ to $t$ here. This is in line with the ultra-relativistic limit that we have taken in the derivation of the wave equations themselves, as explained in Appendix~\ref{Appendix:AxionPhotonEquations}, and henceforth, we will use them interchangeably according to the context. An analogous correspondence is also usually made in neutrino flavour oscillations.}
\begin{equation}
    i \hbar \partial_t|\Psi(t)\rangle = \hat{H}|\Psi(t)\rangle \; ,
\end{equation}
with the effective Hamiltonian above having the same form as the mixing matrix in Eq.\,\eqref{eq:hamiltonian} (also note the equivalence of the Schr\"odinger-like  Eq.\,\eqref{eq:schrodingereq} to the one here). More specifically, in terms of the earlier defined physical parameters, the matrix form of the Hamiltonian operator in Eq.\,\eqref{eq:effhamiltonian} again identifies $\alpha= (m_{\gamma,\parallel}^2 - m_a^2)/(4\hbar^2 \omega) $ and $\beta=g_{a\gamma\gamma} B_{\text{ext}}/2$. In particle mixing scenarios, the equivalence may be modified due to higher-order corrections and vacuum condensate structures\,\cite{Ikeda:2025qac, Binger:1999nj, Blasone:2001du, Capolupo:2019xyd}. Also, this characterisation as a two-state system only holds when Faraday rotation effects are subdominant, as this implies that only the parallel component photon mode will be relevant in axion--photon mixing.

There is a close analogy between the above system and the case of a single photon encountering a beam splitter, with the external magnetic field playing the role of the beam splitter in the above configuration. In both cases, the single quanta (photon or axion) is placed in an entangled state of two modes. We will expand on the latter in the next section while describing quantum mode entanglement in the axion--photon system in its quantum regime.

Before we move on, we would also like to elaborate on a few transparent but interesting features of the quantum regime equations above. First, we note that the oscillation equations that arise in the quantum regime are similar in form to those of the classical theory. One way to view why this is reasonable is from the explicit quantum field-theoretic treatment of the transition probability\,\cite{Ikeda:2025qac}, where the 
classical transition probability emerges as the leading-order result. A complementary, but more direct perspective is structural---in the classical 
treatment, the axion and photon fields satisfy second-order coupled equations of motion (refer to Appendix\,\ref{Appendix:AxionPhotonEquations}) which reduce to first-order 
linear mixing equations upon making a wave ansatz requiring both fields to share a common frequency $\omega$ and using the ultra-relativistic limit---two conditions that together select a single momentum mode (and also lead to the equivalence between using $z$ and $t$). The quantum description, restricted to this same single mode and to the single-excitation sector, is therefore set up to reproduce these equations precisely, and one can recover the wave equation by simply projecting the Schr\"odinger equation onto $\langle z|$. It is therefore not surprising that the classical field amplitude and the single-quanta probability amplitude satisfy equations of the same form; the approximations made on each side are in correspondence and thus, consistent.

It might also be suggestive to consider the magnitude of the field modes ($a, A_\parallel$) and the modulus squared of the corresponding state vectors ($\ket{a}, \ket{\gamma}$). Although the mathematical form of equations is preserved across both regimes, the physical interpretation differs. In the low-occupancy limit ($N_\gamma \sim \mathcal{O}(1)$), it 
represents a probability density in the spirit of the Born rule, whereas in the classical large-occupancy limit ($N_\gamma \rightarrow \infty$) it corresponds to an intensity or energy density associated with classical field waves carrying well-defined phase and amplitude relations. This correspondence or complementarity of amplitudes of individual quanta and their relation to a classical wave might be familiar, for instance, when invoked to describe the double-slit experiment; we can interpret the amplitudes as the probability of detecting a single photon at a detector or the classical interference pattern on a screen\,\cite{Glauber:1963tx, Barnett:2022gzx}. Which interpretation is more appropriate depends on the occupation number of the field, and on an operational level, on whether the measurement apparatus is sensitive to individual 
quanta\,\cite{Barnett:2022gzx, EisamanFanMigdallPolyakov2011, CouteauEtAl2023CommComp, AgarwalTara1991, ZavattaVicianiBellini2004}.

The axion--photon system above is comparable in structure to neutrino oscillations. For vacuum neutrino oscillations in the two-flavour limit, the analogous effective Hamiltonian operator, after dropping a common phase, is
\begin{equation}\label{eq:HeffNeu}
    H_\nu = \frac{\Delta m^2_{ji}}{4 E}\begin{pmatrix}
    -\cos 2\theta_V & \sin2\theta_V \\
    \sin 2\theta_V & \cos 2\theta_V
    \end{pmatrix}\;,
\end{equation}
and the corresponding vacuum oscillation probability is\footnote{`$f$' and `$g$' are labels for the putative neutrino flavours in the interaction or flavour basis}
\begin{equation}
P_{\nu_{f} \rightarrow \nu_{g}}(t) = \sin^2(2\theta_V)\sin^2{\left(\frac{\Delta m_{ji}^2 t}{4 \hbar E}\right)} \; ,
\end{equation}
as is well-known. We have again temporarily restored all the $\hbar$ factors in the above expressions for clarity, while still keeping $c=1$. Note that unlike the axion--photon case, Eq.\,\eqref{eq:effhamiltonian}, the effective Hamiltonian above only contains an $\hbar$ factor implicitly in $E=\hbar\, \omega_{\vec{p}}$. $\theta_V$ is the vacuum oscillation angle relevant to the effective two-flavour system under consideration---for instance, $\theta_{23}$ for atmospheric oscillations, or $\theta_{12}$ for the vacuum contribution in solar neutrino oscillations. Again, we will express the relevant quantum information measures in the neutrino case in terms of the physical parameters $\Delta m_{ji}^2$ and $\theta_V$.

It is important to note that while structurally similar naively from the perspective of a two-level quantum system, axion--photon oscillations are fundamentally different from neutrino oscillations in terms of physical aspects and their actual dependence on the physical parameters of relevance. Thus, although we may map corresponding quantities by defining fictitious quantities in the axion--photon case---for instance, a fictitious mixing angle ($\theta$) through $\sin 2\theta=\beta/\sqrt{\alpha^2+\beta^2}$, and a fictitious mass-squared difference ($\Delta m^2$) through $\Delta m^2/4 E=\sqrt{\alpha^2+\beta^2}$ (see also Eq.\,\eqref{eq:almbdatheta})---the actual dependence on the pertinent physical parameters ($B_{\text{ext}}$, $g_{a\gamma\gamma}$, $m^2_a$ and $m^2_{\gamma,\parallel}$) is obscured by doing so. 

Moreover, the nature of the mixing and the entailing physics is also distinct, particularly in the context of the quantum information measures. For instance, axion--photon oscillations persist classically as a wave phenomenon, arising from linear coupling between two bosonic fields. This is also reflected in how $\hbar$ factors appear in Eqs.\,\eqref{eq:effhamiltonian} and\,\eqref{eq:HeffNeu}. For bosons, in the high occupancy number regime, a coherent state gives a non-zero c-number expectation value of the field operator, tying bosonic coherent states to a corresponding classical field and a classical limit. The quantum regime is alternatively obtained by considering the low-occupancy number limits for the bosons, like in a single-photon experiment. This is unlike neutrino oscillations, which, due to their fermionic nature, do not have a high-occupancy number classical limit in the same sense as above---for fermionic coherent states, the expectation value of the field operator is Grassmann-valued rather than c-number valued, and does not represent a classical field in the same way as that for bosons. 

In the next section, we proceed to discuss various quantum information-related aspects of axion--photon mixing in the relevant quantum regime, where, for instance, a single-photon experiment is being performed in the presence of a large background magnetic field.

\section{Quantum information measures in axion--photon oscillations}
\label{sec:axionphotonQuantumMeasures}

Quantum entanglement may be viewed as a property of composite quantum systems, defined relative to a tensor-product decomposition of the Hilbert space. For a bipartite system $\mathcal{H}=\mathcal{H}_A\otimes\mathcal{H}_B$, a pure state $\ket{\Psi}\in\mathcal{H}$ is separable iff it factorizes as $\ket{\Psi}=\ket{\psi}_A\otimes\ket{\psi}_B$; otherwise $\ket{\Psi}$ is entangled\,\cite{Nielsen:2012yss, Horodecki:2009zz, Plenio:2007zz}. 

A precise characterisation is furnished by the Schmidt decomposition. Consider orthonormal bases $\{\ket{i}_A\}$ and $\{\ket{i}_B\}$ such that
\begin{equation}
\ket{\Psi}=\sum_{i=1}^{r}\sqrt{\lambda_i}\,\ket{i}_A\otimes \ket{i}_B \;,
\qquad
\lambda_i\ge 0 \;,
\qquad
\sum_{i=1}^{r}\lambda_i=1 \; ,
\label{eq:schmidt}
\end{equation}
 where $r$ is the Schmidt rank and $\lambda_i$ are non-negative real numbers. The state is separable iff $r=1$ (equivalently, exactly one Schmidt weight $\lambda_i$ equals $1$). Entanglement then corresponds to $r>1$. For mixed states $\rho_{AB}$ on $\mathcal{H}_A\otimes\mathcal{H}_B$, separability is defined by convex decomposability into product states,
\begin{equation}
\rho_{AB}=\sum_k p_k\,\rho_A^{(k)}\otimes\rho_B^{(k)} \; ,
\qquad
p_k\ge 0 \; ,\quad \sum_k p_k=1 \; ,
\label{eq:mixed-separable}
\end{equation}
and entanglement corresponds to the failure of such a representation, that is, correlations that cannot be modelled as a classical mixture of independent subsystem states from $A$ and $B$\,\cite{PhysRevA.40.4277}. 

To connect this with our earlier discussion, if one starts with an initial photon state in the large-occupancy limit, which is then in the classical regime and can be treated robustly using the classical axion--photon oscillation equations, it may be described by a photon coherent state $\ket{A(0)}_\gamma$
\begin{equation}
\ket{\psi(0)}=\ket{A(0)}_\gamma \otimes \ket{0}_a\;.
\end{equation}
Then, for the bilinear, number-conserving Hamiltonian of Eq.\,\eqref{eq:effhamiltonian} describing the system, the time-evolved state will also be a product of coherent states (see Appendix\,\ref{Appendix:AxionPhotonEquations})
\begin{equation}
\ket{\psi(t)}=\ket{A(t)}_\gamma\otimes \ket{B(t)}_a \;.
\end{equation}
This is a tensor-product state in a factorised form, which implies that in this classical regime, no quantum entanglement will be generated under time evolution, as expected. This motivates, as already mentioned, our consideration of the low-occupancy number regime Fock states (in our case, the single-excitation sector) for the subsequent analysis. We hasten to add that even with coherent states, there have been attempts to generalise the conventional notion of entanglement by defining so-called entangled superpositions of coherent states\,\cite{Sanders1992EntangledCoherentStates}.

In this work, we will focus on pure states in the quantum regime, where notions of quantum entanglement and other quantum information observables are useful descriptions for the axion--photon mixing system. A familiar quantitative diagnostic for bipartite entanglement in a pure state $\ket{\Psi}$ is the entanglement entropy\,\cite{vonNeumann1955, Bennett:1995tk}, defined from the reduced density matrix
$
\rho_A=\Tr_B\!\left(\ket{\Psi}\bra{\Psi}\right)
$
(and similarly $\rho_B$). The von Neumann entropy\;\cite{Bennett:1996gf}
\begin{equation}
S(\rho_A)=-\Tr\!\left(\rho_A\log\rho_A\right) \;,
\label{eq:ent-entropy-def}
\end{equation}
satisfies $S(\rho_A)=S(\rho_B)$ for global pure states and vanishes iff $\ket{\Psi}$ is separable. In the Schmidt basis \eqref{eq:schmidt}, one has $\rho_A=\sum_i \lambda_i \ket{i}_A\bra{i}_A$, and therefore
\begin{equation}
S(\rho_A)=-\sum_{i=1}^{r}\lambda_i\log\lambda_i \;,
\label{eq:ent-entropy-schmidt}
\end{equation}
so that entanglement is encoded in the nontrivial spectrum $\{\lambda_i\}$. For example, the maximally entangled two-qubit Bell state $\ket{\Phi^+}=(\ket{00}+\ket{11})/\sqrt{2}$ yields $\rho_A=\tfrac{1}{2}\mathbb{1}$ and $S(\rho_A)=\log 2$ (and by virtue of its structure, identically for $\rho_B$), which is maximal for a two-dimensional subsystem. Entanglement between two particles, for instance, where the two particles are entangled with respect to some attribute, like their spin, is relatively familiar. In these cases, the subsystems correspond to the first particle (subsystem A) and the second particle (subsystem B). 

\subsection*{Mode entanglement}
In mode entanglement\,\cite{PhysRevA.72.064306, Zanardi:2001zz, Zanardi:2002pwt}, the relevant subsystems are not the particles but field modes (spatial, momentum, polarisation, flavour, etc). The key point is that entanglement here is defined with respect to the mode factorisation $\mathcal{H}=\mathcal{H}_{\tilde{A}}\otimes \mathcal{H}_{\tilde{B}}$ (or its multipartite generalisation). Here, $\tilde{A}$ and $\tilde{B}$ subsystems correspond to some relevant modes, like flavour. An example, relevant for our study, is a delocalised state of the form
\begin{equation}
\ket{\Psi}=\frac{1}{\sqrt{2}}\Bigl(\ket{1}_{\tilde{A}}\ket{0}_{\tilde{B}}+\ket{0}_{\tilde{A}}\ket{1}_{\tilde{B}}\Bigr) \;,
\label{eq:single-excitation-mode-ent}
\end{equation}
which is manifestly entangled across the mode bipartition (i.e., across mode subsystems $\tilde{A}$ and  $\tilde{B}$ now) even though it contains only one particle; tracing out the $\tilde{B}$ mode subsystem gives $ \rho_{\tilde{A}}=\frac{1}{2}\bigl(\ket{1}_{\tilde{A}}\bra{1}_{\tilde{A}}+\ket{0}_{\tilde{A}}\bra{0}_{\tilde{A}}\bigr)$, and hence $S(\rho_{\tilde{A}})=\log 2$. This kind of mode entanglement is familiar in quantum optics as single-photon entanglement\,\cite{Tan:1991oqb, Yurke:1992qpa, PhysRevA.72.064306, Fuwa:2014cra, Guerreiro:2016ytu} where, for instance, a single photon incident on a beam splitter generates an entangled state between two output modes corresponding to transmission ($T$) and reflection ($R$) channels\,\cite{PhysRevLett.92.180401}
\begin{equation}
|1\rangle_T|0\rangle_R \xrightarrow{\text{\tiny{Beam splitter}}} \frac{1}{\sqrt{2}}\left[|1\rangle_T|0\rangle_R+|0\rangle_T|1\rangle_R\right]\; .
\end{equation}
The latter state is equivalent to a maximally entangled Bell state. The axion--photon system is conceptually similar; the external magnetic field induces a mode conversion for a single excitation between an axionic mode and a photon mode, and the distance-dependent superposition generated by the mixing Hamiltonian naturally carries this mode entanglement. As we mentioned in the previous section, the external magnetic field plays the role of the beam splitter in our case; the initial photon, after it encounters the transverse magnetic field, is put in a mode entangled state
\begin{equation}
 \left|1_\gamma,0_a\right\rangle \xrightarrow{~~~\vec{B}_{\text{ext}}~~~}c_1(z)\left|1_\gamma,0_a\right\rangle + c_2(z)\left|0_\gamma,1_a\right\rangle \; .
\end{equation}
The important conceptual takeaway is that nothing in the discussion requires the presence of more than one particle; what is required is a non-trivial tensor-product decomposition of the Hilbert space across the relevant modes in the system, together with a coherent superposition across that decomposition.

In this sense, single-particle entanglement refers to entanglement between modes, rather than between particles\,\cite{PhysRevA.64.042106, PhysRevA.72.064306}. This aspect, and some of the ensuing entanglement properties have been studied in the literature for neutrino oscillations\,\cite{Blasone:2007wp, Blasone:2007vw}. Here, the neutrino interaction or flavour states ($\ket{\nu_\alpha}$) are a coherent superposition of propagation or mass eigenmodes ($\ket{\nu_i}$),
\begin{equation}
\ket{\nu_\alpha}=\sum_i U_{\alpha i}\ket{\nu_i} \; ,
\qquad
\ket{\nu_\alpha(t)}=\sum_i U_{\alpha i}\,e^{-iE_i t}\,\ket{\nu_i}\; .
\label{eq:nu-superposition}
\end{equation}
This furnishes a physically motivated partition where mode entanglement may exist and can be studied\,\cite{Blasone:2007wp, Blasone:2007vw}.

In general, given some physically motivated partition into modes, one may construct reduced density matrices by partial tracing, and quantifying the resulting mode entanglement through Eqs.\,\eqref{eq:ent-entropy-def} and \eqref{eq:ent-entropy-schmidt}. While entanglement is invariant under local unitaries\footnote{A local unitary is a unitary operator that factorises with respect to the subsystem split} within a fixed subsystem split, its presence and physical interpretation are intrinsically tied to the choice of degrees of freedom that define the particular tensor-product structure. Thus, mode entanglement potentially provides a natural language for quantifying single-particle entanglement phenomena in field-theoretic settings.

We are primarily interested in the axion--photon system and its mixing in the quantum regime, as espoused already in the previous discussions. The regime of interest is the single-excitation sector of axions and photons in the presence of an external magnetic field, where the classical oscillation aspects do not furnish the appropriate description, and one has to treat the state conversions quantum mechanically. As we already mentioned, among the motivations for focusing on the quantum regime and the single-excitation sector are recent studies on the use of single-photon experiments in the context of quantum-enhanced searches of axions\,\cite{Dixit:2020ymh,Yang:2022uil,Devlin:2026czx} and general questions regarding the quantum aspects that may be present in the axion--photon system\,\cite{Zheng:2024kxn,Bao:2025nsd}.

Exactly the same operational techniques as in the neutrino system \cite{Blasone:2007wp, Blasone:2007vw} apply to the axion--photon system, except that the physically relevant basis may be chosen either as the interaction basis $\{\ket{\gamma},\ket{a}\}$ or as the propagation basis $\{\ket{\tilde a_1},\ket{\tilde a_2}\}$. Because these two bases are related by a rotation that mixes the subsystems, entanglement and coherence measures are not numerically identical in the two descriptions. This is not a contradiction. Entanglement is invariant only under local unitaries with respect to a fixed tensor-product decomposition, whereas the map from interaction modes to propagation modes changes the very mode partition with respect to which the bipartite structure is defined\,\cite{Zanardi:2001zz}. We will therefore study quantum information measures in both these bases to gain complementary perspectives for the axion--photon system. However, we emphasise here that in practice, photon counting statistics are most naturally extracted in the interaction basis---it is photons that are detected in any realistic experiment and not these propagating modes. The propagation basis analysis is therefore presented as a theoretical complement; it is a useful perspective to have for the intrinsic mixing structure of the system and provides an independent consistency check (for instance, limits of the mixing parameters $\alpha$ and $\beta$), even if it is one step removed from experimentally inferrable quantities.

Before we move on to the calculations, we note some numerical estimates of the mixing parameters $\alpha$ and $\beta$, specifically in this context, since all our calculations henceforth will naturally involve these parameters and it is worth getting an idea of the numerical scales involved (especially for axion searches involving quantum sensing). For $B_{\text{ext}}\sim 10\,\mathrm{T}$ and $g_{a\gamma\gamma}\sim\,10^{-13}\,\mathrm{GeV}^{-1}$, we have $\beta \sim 10^{-19}\; \mathrm{eV}$. Note that the experimental bounds on $g_{a\gamma\gamma}$ are usually quoted as exclusion regions in the $(m_a, g_{a\gamma\gamma})$ 
plane\,\cite{Irastorza:2018dyq,DiLuzio:2020wdo}, so using $g_{a\gamma\gamma} \sim 10^{-13}\,\mathrm{GeV}^{-1}$ is an order-of-magnitude representation, not a mass-specific bound. Estimating $\alpha$ independently is trickier---in axion search experiments, the effective photon mass $m_{\gamma,\parallel}$ is tuned to match a target axion mass that is of interest to the search, and thus, the experiment is deliberately driven towards the resonance condition $\alpha \sim 0$. In this situation, $\beta$ dominates the mixing, and the transition probability is maximal. Nevertheless, if we consider the case where $\lvert \alpha \rvert$ is at least of the same order of magnitude as $\beta \sim 10^{-19}\;\textrm{eV}$ and there is no effective photon mass $m_{\gamma,\parallel} = 0$, for searches using microwave photons ($\omega \sim 10^{-5}\;\textrm{eV}$) we have an approximate axion mass corresponding to $m_a \sim 10^{-12}\;\textrm{eV}$.

\subsection*{Mode decomposition in the axion--photon system}
For the axion--photon problem, the physically relevant tensor-product structure in the interaction basis is the mode decomposition represented by the Hilbert-space factorisation
\begin{equation}
\mathcal H=\Hg\otimes\Ha \;,
\label{eq:mode-factorization}
\end{equation}
where $\Hg$ is the Fock space of the photon mode polarised parallel to the external magnetic field and $\Ha$ is the Fock space of the axion mode. In the single-excitation sector regime discussed earlier, the dynamics can be understood in the effective Hilbert space
\begin{equation}
\mathcal H_{1\text{-exc}}=\mathrm{span}\!\left\{\ket{1_\gamma,0_a},\ket{0_\gamma,1_a}\right\} \;.
\label{eq:single-excitation-sector}
\end{equation}
Introducing bosonic creation operators $\hat{b}^\dagger$ and $\hat{c}^\dagger$ for the photon and axion modes, as in the previous section, one has
\begin{equation}
\ket{\gamma}\equiv\ket{1_\gamma,0_a}=\hat{b}^\dagger\ket{0_\gamma,0_a},
\qquad
\ket{a}\equiv\ket{0_\gamma,1_a}=\hat{c}^\dagger\ket{0_\gamma,0_a} \;.
\label{eq:fock-basis-defs}
\end{equation}
The two-dimensional oscillation problem is therefore not merely analogous to a qubit system. Rather, it is the restriction of the full bosonic Fock space to a single-mode single-excitation subspace.

In the interaction (flavour) basis of $|\gamma \rangle$ and $|a\rangle$, the state evolves along the propagation direction $z$, and for a pure photon initial state $|\gamma \rangle$
\begin{equation}
|\Psi(0)\rangle = |\gamma \rangle \equiv \ket{1_\gamma,0_a}\; ,
\end{equation}
the evolved state in the interaction basis takes the form  
\begin{equation}\label{eq:psiatz}
    |\Psi(z)\rangle = c_1(z) |\gamma\rangle + c_2(z) |a\rangle \equiv c_1(z)\ket{1_\gamma,0_a}+c_2(z)\ket{0_\gamma,1_a} \; .
\end{equation}
which is suggestive of the presence of (mode) entanglement due to a naturally motivated tensor product factorisation of the Hilbert space, and consequently, the basis vectors. The coefficients $c_1(z)$ and $c_2(z)$ derived from the dynamical evolution equation are given by
\begin{equation}\label{eq:c1c2coeff}
    \begin{aligned}
    c_1(z) &= \cos^2\theta\, e^{-i\lambda_+ z} + \sin^2\theta\, e^{-i\lambda_- z} \; ,\\
    c_2(z) &= \sin\theta \cos\theta \left(e^{-i\lambda_+ z} - e^{-i\lambda_- z}\right) \;,
    \end{aligned}
\end{equation}
with
\begin{equation}\label{eq:almbdatheta}
    \begin{aligned}
        \lambda_\pm &= \pm\sqrt{\alpha^2 + \beta^2} \;,\\
        \sin(2\theta) &=\frac{\beta}{\sqrt{\alpha^2+\beta^2}} \;.
    \end{aligned}
\end{equation}
Note that $c_1(z)$ and $c_2(z)$ are subject to the normalisation condition $|c_1(z)|^2 + |c_2(z)|^2 = 1$ and are related to the physical probabilities as 
\begin{equation}\label{eq:probabilityc1c2}
    \begin{aligned}
        |c_2(z)|^2=P_{\gamma\to a}(z)&=\frac{\beta^2}{\alpha^2+\beta^2}\sin^2(z\sqrt{\alpha^2+\beta^2}) \;,\\
        |c_1(z)|^2&=1-P_{\gamma\to a}(z) \;.
    \end{aligned}
\end{equation}
Defining,
\begin{equation}
\rho^{(a)} = |\Psi(z)\rangle \langle \Psi(z)| \; ,
\end{equation}
and taking the partial trace with respect to the axion subsystem, we have the reduced density matrix
\begin{equation}
\rho_{\gamma}^{(a)} = \text{Tr}_a[\rho^{(a)}] = |c_1(z)|^2 |1_\gamma\rangle \langle 1_\gamma| + |c_2(z)|^2 |0_\gamma\rangle \langle 0_\gamma| \; .
\end{equation}

\subsection{Entanglement entropy}
\subsubsection*{von Neumann entropy}
\begin{figure}[ht!]
    \centering
    \begin{subfigure}[b]{0.75\textwidth}
        \centering
        \includegraphics[width=\linewidth]{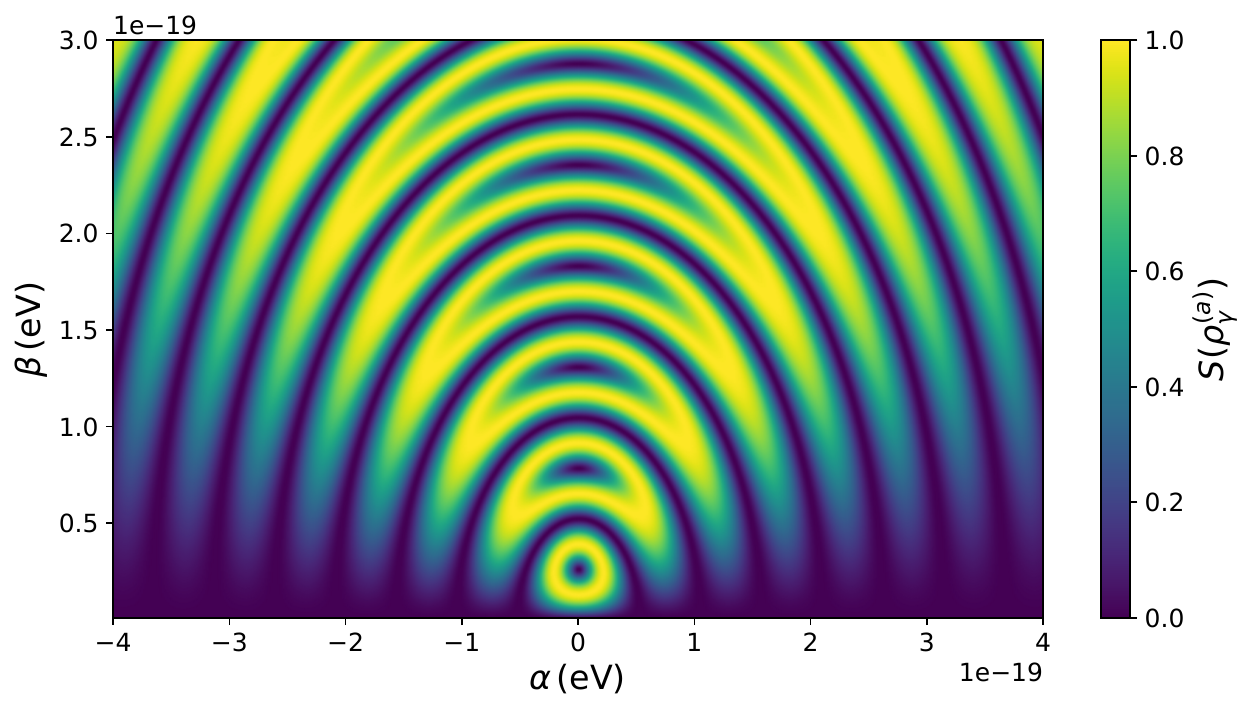}
        \label{fig:VN_interaction}
    \end{subfigure}
    \hfill
    \begin{subfigure}[b]{0.75\textwidth}
        \centering
        \includegraphics[width=\linewidth]{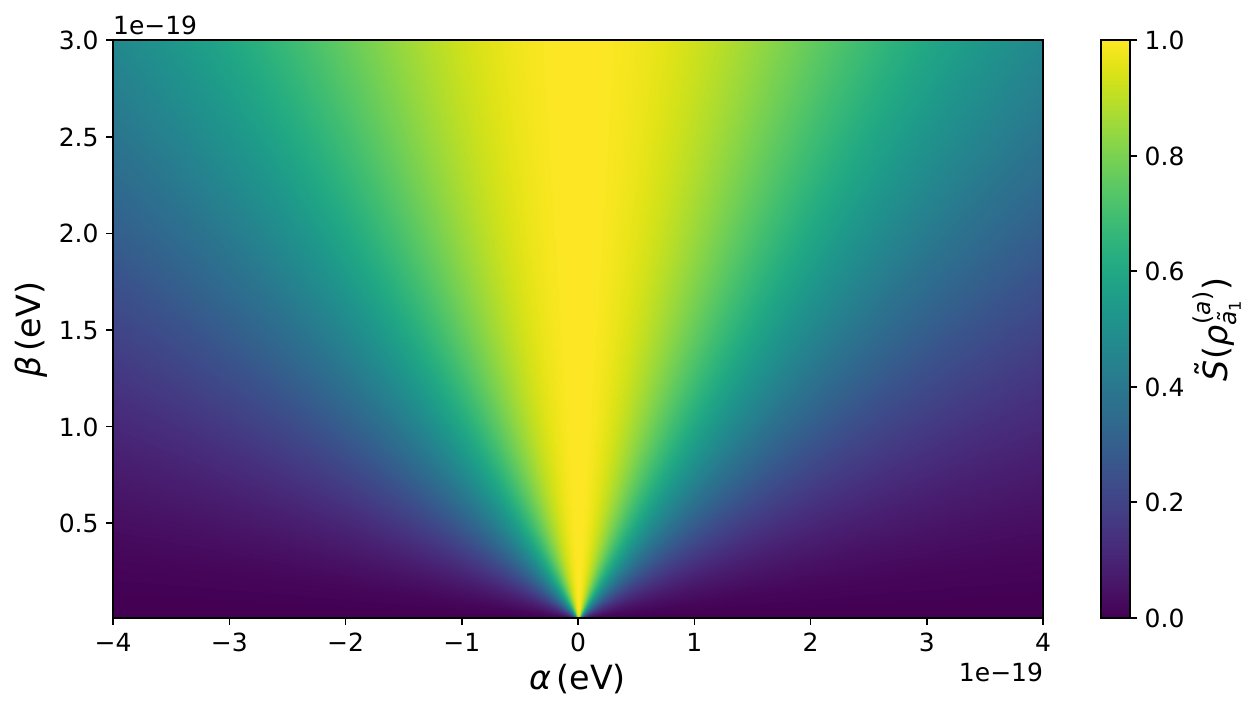}
        \label{fig:VN_propagation}
    \end{subfigure}
    \caption{An illustrative plot of the von Neumann entropy for the photon subsystem in the interaction and propagation basis, plotted as a function of the system parameters $\alpha$ and $\beta$. The interaction basis entropy is calculated at a fixed $z =6 \times 10^{19} \, \mathrm{eV}^{-1}$ (as for the transition probability in Fig.~\ref{fig:transitionprobability}). Notice how the $z$-dependence in the interaction basis changes the behaviour of the entropy in comparison to the static behaviour present in the propagation basis, we observe that it is modulated by the sinusoidal behaviour characteristic of the transition probability $P_{  \gamma\rightarrow a}(z)$.}
    \label{fig:VN_entropy}
\end{figure}

The von Neumann entropy is given by
\begin{equation}
    S(\rho_{\gamma}^{(a)}) = -\operatorname{Tr}\left(\rho_{\gamma}^{(a)} \log_2\rho_{\gamma}^{(a)}\right) \;.
\end{equation}
For the axion--photon oscillations, the von Neumann entropy in the interaction basis is calculated to be
\begin{align}\label{eq:EEap}
    S(\rho_{\gamma}^{(a)}) & = -\left[ 
        P_{\gamma \rightarrow a}(z) \log_2 P_{\gamma \rightarrow a}(z)
        + \lbrace 1 - P_{\gamma \rightarrow a}(z)\rbrace \log_2\lbrace 1 - P_{\gamma \rightarrow a}(z)\rbrace
    \right] \nonumber \\
    & \equiv - \left[\frac{\beta^2}{\alpha^2+\beta^2} \sin^2\left(z \sqrt{\alpha^2+\beta^2} \right) \log_2\left( \frac{\beta^2}{\alpha^2+\beta^2} \sin^2\left(z \sqrt{\alpha^2+\beta^2} \right) \right) \right. \nonumber \\
    &\left. + \left( 1 - \frac{\beta^2}{\alpha^2+\beta^2} \sin^2\left( z \sqrt{\alpha^2+\beta^2}  \right) \right)
    \log_2\left( 1 - \frac{\beta^2}{\alpha^2+\beta^2} \sin^2\left( z \sqrt{\alpha^2+\beta^2} \right) \right)
    \right] \; .
\end{align}
Although this can be interpreted as the binary Shannon entropy $H_2(P_{\gamma\to a})$ evaluated on the conversion probability, here we emphasise that the appearance of $P_{\gamma \rightarrow a}(z)$ in the above equation is strictly because of the probability amplitudes $c_1(z)$ and $c_2(z)$ of the single-excitation sector (Eq.\,\eqref{eq:psiatz}), and thus is interpreted as a quantum entanglement. A similar perspective will also apply to other measures. Eq.\,\eqref{eq:EEap} vanishes when $P_{\gamma\to a}$ vanishes and when $P_{\gamma\to a}$ approaches unity, both cases which correspond to the state being separable in the interaction basis. Fig.\,\ref{fig:VN_entropy} (top) illustrates how the von Neumann entropy in the interaction basis behaves as a function of the parameters $\alpha$ and $\beta$. 

The global maximal value of $S(\rho_{\gamma}^{(a)})$ is attained when $P_{\gamma\to a}=1/2$, which also corresponds to the reduced state being maximally mixed. Thus, maximal entanglement in the interaction basis is attained, as a function of $\alpha$ and $\beta$, when the conversion dynamics produces an equal-weight superposition of axion and photon flavour modes. Strictly speaking, there is in general a class of solutions which are dependent on $z$ (Fig.\,\ref{fig:VN_entropy}, top)---for instance, the resonance condition $\alpha=0$, together with propagation to a point where $\sin^2(\beta z)=1/2$ (modulo the periodicity). In contrast, for some given, fixed $\alpha$ and $\beta$, we would have somewhere along the path
\begin{equation}\label{eq:maxent}
 S(\rho_{\gamma}^{(a)})^{\,\max }=\left\{\begin{array}{lr}
1, ~~& \beta^2 \geq \alpha^2 \; , \\
-\frac{\beta^2}{\alpha^2+\beta^2} \log _2\left(\frac{\beta^2}{\alpha^2+\beta^2}\right)-\frac{\alpha^2}{\alpha^2+\beta^2} \log _2\left(\frac{\alpha^2}{\alpha^2+\beta^2}\right), ~~& \beta^2 < \alpha^2 \; .
\end{array}\right.
\end{equation}

In terms of the true external physical parameters, we see that the behaviour of the constants $\alpha$ and $\beta$, and their subsequent influence on the entropy is an interplay between the relative strengths of the masses $m_{\gamma, \parallel}$ and $m_a$, the coupling constant $g_{a\gamma\gamma}$, the magnetic background $B_{\text{ext}}$, and to a lesser extent, the energy of the mode selected, represented by $\omega$ (refer to Eq.\,\eqref{eq:alphabeta} for the precise relations). We observed that the complete maximisation of the entropy may occur at the resonance condition $\alpha = 0$, which is equivalent to $m^2_{\gamma, \parallel} = m^2_a$. Other scenarios exist where the entropy, viewed as functions of $\alpha$ and $\beta$, may approach its maximal value asymptotically. This occurs when we consider some combination of the following---a high energy single-excitation sector where $\omega$ is much larger than the mass-squared difference $\Delta m_{\gamma a}^2=m^2_{\gamma, \parallel} - m^2_a$, or when the product $g_{a\gamma\gamma}B_{\text{ext}}$ is much larger than the aforementioned mass-squared difference (i.e. $\beta\gg\alpha$). Since the coupling constant $g_{a\gamma\gamma}$ is known to have an upper bound\;\cite{Caputo:2024oqc}, it is really the strength of the magnetic field that drives this limit for the saturation of the entropy value. Note that these maximisation conditions also maximise the transition probability attainable somewhere along the path, and thus may be naturally suited to the conditions of experimental axion searches.

\subsubsection*{Linear entropy}
Next, we look at the linear entropy\,\cite{Horodecki:2009zz}, which is derived from the von Neumann entropy and is given by
\begin{equation}
    S_L^{(\gamma;a)}(\rho^{(a)}) = 2\left(1 - \operatorname{Tr}_{\gamma}\left[\left(\rho_{\gamma}^{(a)}\right)^2\right]\right) \;.
\end{equation}
It is the same for the photon subsystem and the axion subsystem (i.e. $S_L^{(\gamma;a)}(\rho^{(a)}) = S_L^{(a;\gamma)}(\rho^{(\gamma)}) $) and may be calculated as
\begin{align}
S_L^{(\gamma;a)}(\rho^{(a)}) & = 4 P_{a\to\gamma}(z) \left( 1- P_{a\to\gamma}(z)\right) \nonumber \\
& \equiv 4 \left( \frac{\beta^2}{\alpha^2+\beta^2}\sin^2(z\sqrt{\alpha^2+\beta^2}) \right) \left(1 - \frac{\beta^2}{\alpha^2+\beta^2}\sin^2(z\sqrt{\alpha^2+\beta^2})\right) \;.
\end{align}
The linear entropy naturally inherits the behaviour of the entanglement entropy with respect to $\alpha$ and $\beta$, and thus, the behaviour of the linear entropy with respect to the external parameters follows directly from the discussion on the trends in the von Neumann entropy above. It is, however, also simpler to observe the features of this measure for the maximisation conditions. For instance, in the near-resonance limit of $\alpha \sim 0$ and when $\beta\gg\alpha$, we observe that $S_L^{(\gamma;a)}(\rho^{(a)}) \sim \sin^2(2z\sqrt{\alpha^2 + \beta^2})$, and thus, the characteristic period of the (linear) entropy over $z$ is half of that of the transition probability (Figs.\,\ref{fig:transitionprobability} and \ref{fig:VN_entropy}).

While it is not the case for our simpler bipartite qubit scenario, the complexity of calculating the linear entropy (which is the linearised approximation to the von Neumann entropy) scales better than von Neumann entropy with the dimension of the Hilbert space; one simply needs the trace of the density matrix squared (also known as the purity) rather than the full eigenspectrum, and so, the linear entropy is likely to be a preferred estimate of the entanglement entropy for higher dimensions. A potential extension of this analysis for a general multi-mode expansion (which would live in a higher-dimensional Hilbert space) would undoubtedly benefit from a baseline calibration of this measure in this scenario, and it also allows us to compare our results with the neutrino case as discussed in \cite{Blasone:2007wp, Blasone:2007vw}.

As alluded to before, one can also calculate these measures in the propagation basis. Strictly speaking, it is primarily to contrast and attain a complementary perspective of entanglement for axion--photon mixing, as the propagation modes are not the ones that are created or detected. In this basis, the evolution of the state is represented by
\begin{equation}
    |\Psi(z)\rangle = \tilde{c}_1(z) |\tilde{a}_{1}\rangle + \tilde{c}_2(z) |\tilde{a}_{2}\rangle \; ,
\end{equation}
with the coefficients $ \tilde{c}_1(z) $ and $ \tilde{c}_2(z) $ given by
\begin{equation}\label{eq:c1c2coeffprop}
    \begin{aligned}
        \tilde{c}_1(z) &= e^{-i\lambda_+ z}\;\cos\theta \; ,\\
        \tilde{c}_2(z) &= e^{-i\lambda_- z}\;\sin\theta \; ,
    \end{aligned}
\end{equation}
for the specific case of an initially pure photon state. Defining the reduced density matrix after tracing out the $\tilde{a}_2$ basis as
\begin{equation}
\rho_{\tilde{a}_1}^{(a)} = 
\text{Tr}_{\tilde{a}_2}[\rho^{(a)}] = |\tilde{c}_1(z)|^2 |1\rangle \langle 1| +|\tilde{c}_2(z)|^2 |0\rangle \langle 0| \;,
\end{equation}
we find the von Neumann entropy and linear entropy in the propagation basis to be
\begin{align}
    S(\rho_{\tilde{a}_1}^{(a)})=-\frac{1}{2}\left(1-\frac{\alpha}{\sqrt{\alpha^2+\beta^2}}\right)\log_2{\left[\frac{1}{2}\left(1-\frac{\alpha}{\sqrt{\alpha^2+\beta^2}}\right)\right]}-\frac{1}{2}\left(1+\frac{\alpha}{\sqrt{\alpha^2+\beta^2}}\right)\log_2{\left[\frac{1}{2}\left(1+\frac{\alpha}{\sqrt{\alpha^2+\beta^2}}\right)\right]} \; ,
\end{align}
\begin{equation}
S_L^{(\tilde{a}_1;\tilde{a}_2)}(\rho^{(a)}) = \frac{\beta^2}{\alpha^2+\beta^2} \; .
\end{equation}
The $z$-independence of the propagation basis measures is not surprising. The propagation eigenmodes accumulate only phases under evolution, and therefore, the occupation probabilities of those modes remain constant. The interaction basis, by contrast, is the basis in which physically observable axion--photon conversion occurs, so all dynamical oscillatory behaviour is visible there. We observe that the resonance condition ($\alpha = 0$) and the large magnetic background limit ($\beta \gg \alpha$) saturate, the latter in an approximate sense, both the von Neumann, as well as the linear entropy in the propagation basis. The resonant solution $\alpha=0$ is the only exact solution for maximal entanglement in the propagation basis. Fig.\;\ref{fig:VN_entropy} illustrates a comparison of the propagation and interaction basis von Neumann entropies. The dynamical and static natures of the respective entanglement measures are evident from their different dependences on $\alpha$ and $\beta$.

\subsection{Concurrence, PPT criterion and Negativity}
For pure bipartite states of Schmidt rank two, the linear entropy, concurrence, negativity, and entropy of entanglement are all monotone functions of the same pair of non-zero Schmidt coefficients. Consequently, these quantities all encode a similar ordering of entanglement for the present system, although they do so in different functional forms and numerical scales. We calculate and investigate these standard entanglement measures now.
\subsubsection*{Concurrence}
Concurrence\,\cite{Wootters:1997id} is an alternative entanglement monotone, with a different functional dependence on the Schmidt weights compared to entanglement entropy. As mentioned above, for pure two-qubit states, it contains similar entanglement information as the entanglement entropy measure. Nevertheless, for mixed states or higher-dimensional systems, their meanings and applicability may differ substantially. 

In the axion--photon system, in the interaction basis, it may be calculated to be
\begin{align}
    \mathcal{C}(z) &= 2 \sqrt{P_{\gamma\to a}(z) \left( 1-P_{\gamma\to a}(z) \right)}\nonumber \\
    &\equiv  2 \sqrt{\frac{\beta^2}{\alpha^2+\beta^2} \sin^2\left( z \sqrt{\alpha^2+\beta^2}  \right) \left(1 - \frac{\beta^2}{\alpha^2+\beta^2} \sin^2\left( z \sqrt{\alpha^2+\beta^2}  \right)\right)} \;.
\end{align}
In the propagation basis, the analogous quantity is
\begin{equation}\label{eq:concpropbasis}
\tilde{\mathcal{C}} =  \frac{\abs{\beta}}{\sqrt{\alpha ^2 + \beta^2}} \;.
\end{equation}
It has a constant value that is independent of $z$, as expected in the propagation basis. The resonance condition $\alpha = 0$ is again the only exact solution for maximal concurrence here.

We note a few characteristics of concurrence, particularly in the interaction basis, that will also naturally apply to features of other measures related to it. For instance, if we consider the near-resonance ($\alpha \sim 0$) or the high magnetic background limit ($\beta\gg\alpha$) of concurrence, at a fixed $z = \pi/(2\sqrt{\alpha^2 + \beta^2})$, we observe that $\mathcal{C}(z = \pi/(2\sqrt{\alpha^2 + \beta^2})) \sim 2|\alpha|/\beta$. In this regime, the concurrence is proportional to the mass-squared difference $\Delta m_{\gamma a}^2 $, but suppressed by the energy of the photon $\omega$ (as it appears in $\alpha$), and the magnitude of the external magnetic field $B_{\text{ext}}$ (appearing in $\beta$). This suppression has a direct experimental interpretation---a stronger magnetic field not only increases the overall conversion rate $P_{\gamma\to a}$, but it also makes the concurrence less sensitive to small variations from resonance as it absorbs the experimental variance ($\Delta\alpha$) in the resonance parameter $\alpha$. This is useful, as it effectively broadens the ``resonance window'' that can be used and thus, can be a helpful diagnostic to calibrate and quantify the behaviour of concurrence (and other related measures) in the near-resonance limit. Note that this is also precisely the regime targeted when scanning for massive axions in experiments; by tuning the effective photon mass, one obtains the exclusion regions in axion searches\,\cite{CAST:2004gzq,CAST:2007jps,Irastorza:2018dyq,DiLuzio:2020wdo,Caputo:2024oqc}. Notice also how the functional form of the concurrence suppresses the variance in $\alpha$---in the context of axion searches, where the detuning by the effective photon mass has to be extremely precise owing to the size of $\beta$ ($\sim 10^{-19}\;\textrm{eV}$), the concurrence calculated for two values of $\alpha_{\text{exp}}$ which are within an order of $\sim10^2$ of each other will have a concurrence value that only differs by a factor of 10, instead of a factor of $10^2$ as in the linear entropy.

In Fig.\;\ref{fig:QI_measures_comparison}, we make a comparison of the concurrence measure against other useful entanglement measures, as a function of the transition probability $P_{\gamma\to a}(z)$. Note that the actual transition probability is dependent on the physical parameters $\alpha$ and $\beta$, and only spanned between $[0,\,\beta^2/(\alpha^2+\beta^2)]$. It is clear from the figure that when $P^{\,\max}_{\gamma \rightarrow a} < 1/2$, for fixed $\alpha$ and $\beta$, the point where the quantum information measures attain their maximum value coincides with the point where the conversion probability is also at its maximum. On the other hand, for some given, fixed values of $\alpha$ and $\beta$, the maximum value that may be attained somewhere along the path is
\begin{equation}
 \mathcal{C}^{\,\max }= \begin{cases}1, & \beta^2 \geq \alpha^2\;, \\ \frac{2\,\abs{\alpha \beta}}{\alpha^2+\beta^2}, & \beta^2<\alpha^2 \; .\end{cases}
\end{equation}
We note in passing that the above is in general different from the constant value $\tilde{\mathcal{C}}$, calculated in Eq.\,\eqref{eq:concpropbasis} for the propagation basis.

Using the structural similarity with the two-neutrino system, we can calculate the expressions for the concurrence in the interaction and the propagation basis, respectively as
\begin{equation}
\begin{aligned}
    \mathcal{C}(t) = 2 &\sqrt{\sin^2(2\theta_V)\sin^2{\left(\frac{\Delta m_{ji}^2t}{4E}\right)}\left(1-\sin^2(2\theta_V)\sin^2{\left(\frac{\Delta m_{ji}^2t}{4E}\right)}\right)} \; ,\\
    &\qquad\qquad\qquad\qquad \tilde{\mathcal{C}} = \sin \left( 2\theta_V\right) \; .
\end{aligned}
\end{equation}
For general, fixed physical parameters $\theta_V$ and $\Delta m_{ji}^2$, the maximum value of the concurrence measure $\mathcal{C}(t)$ in the neutrino case is $\sin 4 \theta_V$ for $\theta_V < \pi/8$, and $1$ for larger vacuum angles. In the mass basis, we note that the concurrence $\tilde{\mathcal{C}}$, as a function of the physical parameters $\theta_V$ and $\Delta m_{ji}^2$, is maximised at $\theta_V = \pi/4$, corresponding to maximal mixing.

Here, it is worth recalling that the  atmospheric neutrino mixing angle is observed to be $\theta_{23} \approx 49^
\circ$\;\cite{Esteban:2020cvm, T2K:2023smv, NOvA:2021nfi}, placing the atmospheric sector near the entanglement-maximising point of the two-flavour system. These types of coincidences have sometimes been considered suggestive in other areas of high energy physics, and studied in the literature---speculating on whether entanglement extremisation principles, either maximisation or minimisation, potentially shed light on the observed structure of fundamental quantities like, for instance, flavour-mixing matrices\,\cite{Cervera-Lierta:2017tdt, Thaler:2024anb}. We will also discuss later how a similar condition arises from quantum speed limit (QSL) considerations. We do not pursue this speculative direction here, but note that the axion--photon system provides a structurally similar setting, with the resonance condition $\alpha = 0$ and the $\beta\gg\alpha$ saturation limit playing a similar role to the constraint on the angle $\theta_V$.

\subsubsection*{PPT criterion and Negativity}
A general criterion necessary to distinguish between separable and entangled states for bipartite systems is the Positive Partial Transpose (PPT) or Peres-Horodecki criterion\,\cite{Peres:1996dw, Horodecki:1996nc}. It states that the density operator $ \rho_{AB} $ of a bipartite qubit system is separable if and only if its partial transpose over any one subsystem (say B), $ \rho^{T_B} $ has no negative eigenvalues\footnote{Here, we emphasise that although in general it is only a necessary condition, it is necessary and sufficient for the present bipartite qubit case.}. Despite only being necessary and not sufficient for confirming the separability of a general bipartite state, the PPT criterion and the subsequent negativity measures are extremely ubiquitous given their suitability for general mixed states (where the standard Schmidt decomposition does not apply).

Applying the partial transpose over the axion subsystem in the full density matrix, we have\footnote{The bra-ket form of this matrix is $\rho^{(a)T_a}=|c_1(z)|^2\ket{10}\bra{10}+|c_2(z)|^2\ket{01}\bra{01}+c_1(z)c_2^*(z)\ket{00}\bra{11}+c_1^*(z)c_2(z)\ket{11}\bra{00}$ \;.} 
\begin{equation}
    \rho^{(a)T_a} = \begin{pmatrix}
        0 & 0 & 0 & c_1^*(z)c_2(z) \\
        0 & |c_1(z)|^2 & 0 & 0 \\
        0 & 0 & |c_2(z)|^2 & 0 \\
        c_1(z) c_2^*(z) & 0 & 0 & 0
    \end{pmatrix} \;.
\end{equation}
The eigenvalues of this matrix are found to be: {$|c_1(z)|^2$, $|c_2(z)|^2$, $+|c_1(z)||c_2(z)|$, and $-|c_1(z)||c_2(z)|$}. Since we obtain a negative eigenvalue, the state is entangled. In our case, since we have one negative eigenvalue $ -|c_1(z)||c_2(z)| $, the negativity\,\cite{Zyczkowski:1998yd, Vidal:2002zz} is given by
\begin{equation}
    \mathcal{N}(\rho^{(a)})\left[z\right] = |c_1(z)||c_2(z)|=\sqrt{\frac{\beta^2}{\alpha^2+\beta^2} \sin^2\left( z \sqrt{\alpha^2+\beta^2}  \right) \left(1 - \frac{\beta^2}{\alpha^2+\beta^2} \sin^2\left( z \sqrt{\alpha^2+\beta^2}  \right)\right)} \;.
\end{equation}
The logarithmic negativity\,\cite{Plenio:2005cwa} is then defined as
\begin{equation}
    E_{\mathcal{N}}(\rho^{(a)}) = \log_2\left(2 \mathcal{N}(\rho^{(a)}) + 1\right) \;,
\end{equation}
which is calculated to be
\begin{equation}
 E_{\mathcal{N}}(\rho^{(a)})\left[z\right]= \log_2\left(2 \sqrt{\frac{\beta^2}{\alpha^2+\beta^2} \sin^2\left( z \sqrt{\alpha^2+\beta^2}  \right) \left(1 - \frac{\beta^2}{\alpha^2+\beta^2} \sin^2\left( z \sqrt{\alpha^2+\beta^2}  \right)\right)} +1\right) \;.
\end{equation}

\begin{figure}[h!]
    \centering
    \includegraphics[width=0.65\linewidth]{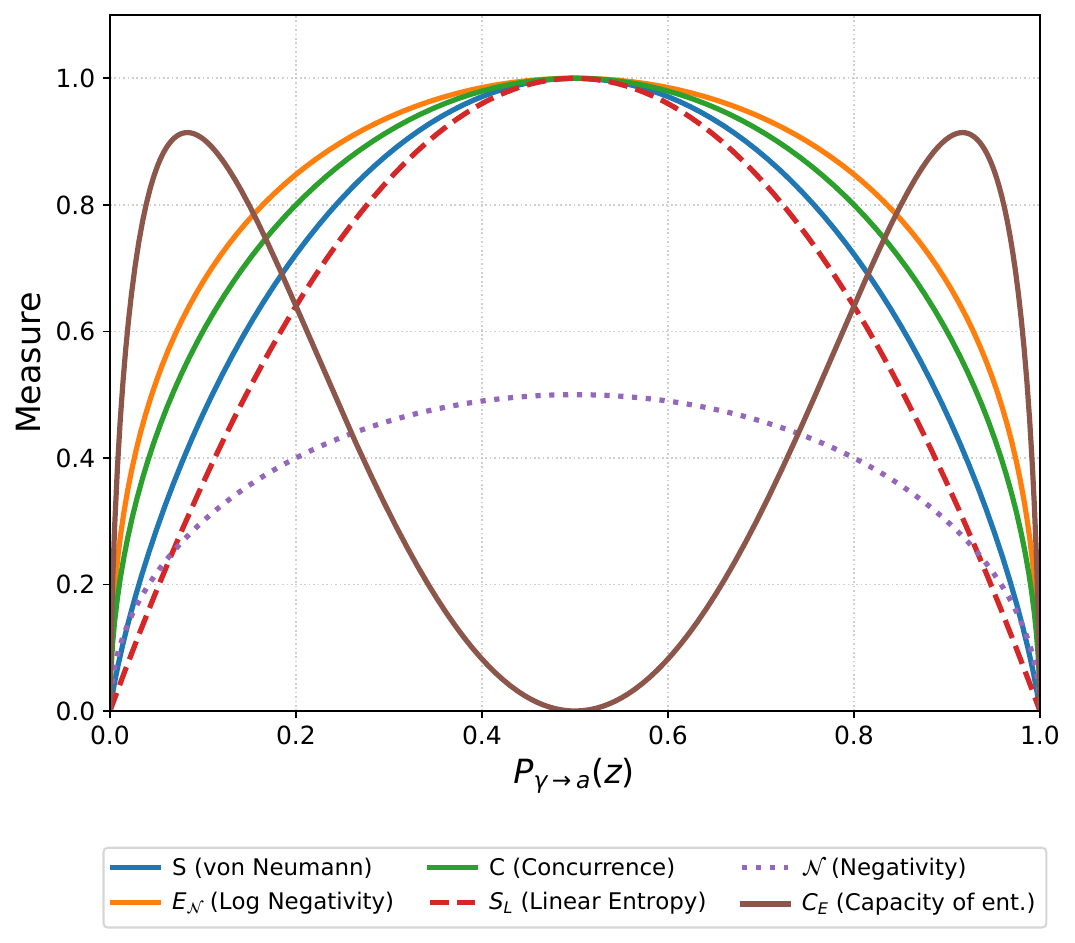}
    \caption{Quantum information measures calculated for the axion--photon system, plotted as functions of the transition probability $P_{\gamma\rightarrow a}(z)$. Notice that all measures, except the capacity of entanglement (which measures the fluctuation rather than the magnitude of the entanglement spectrum), follow a similar evolution---the measures, although providing complementary perspectives, are related to each other as they involve the same Schmidt weight in different functional forms. It is also important to emphasise again that the maximum transition probability attained depends on the values of $\alpha$ and $\beta$, and does not necessarily always span the full range $[0,1]$.}
    \label{fig:QI_measures_comparison}
\end{figure}

Unsurprisingly, we note that some of the same intriguing regimes of $\alpha$ and $\beta$ pointed out earlier, also generate noteworthy values of negativity and logarithmic negativity. Here, the more negative the eigenvalue of the partially transposed matrix, the more entangled the state is, and the threshold of entanglement is captured by values of $\alpha$ and $\beta$ that set the eigenvalue $ -|c_1(z)||c_2(z)| $ to zero (see, Eqs.\,\eqref{eq:c1c2coeff} and \eqref{eq:almbdatheta}). The structure of the expression implies that the $1/\beta$ scaling that we observed in the concurrence case is similarly carried over, and retains its physical implication of suppressing the experimental variance in $\alpha$. The profile of the negativity and logarithmic negativity is again compared to other standard measures in Fig.\;\ref{fig:QI_measures_comparison}.

In the propagation basis analysis, we have
\begin{equation}
    \rho^{(a)T_2} = \begin{pmatrix}
        0 & 0 & 0 & \tilde{c}_1^*(z)\tilde{c}_2(z) \\
        0 & |\tilde{c}_1(z)|^2 & 0 & 0 \\
        0 & 0 & |\tilde{c}_2(z)|^2 & 0 \\
        \tilde{c}_1(z) \tilde{c}_2^*(z) & 0 & 0 & 0
    \end{pmatrix} \;.
\end{equation}
The eigenvalues of this matrix are found to be: {$|\tilde{c}_1(z)|^2$, $|\tilde{c}_2(z)|^2$, $+|\tilde{c}_1(z)||\tilde{c}_2(z)|$, and $-|\tilde{c}_1(z)||\tilde{c}_2(z)|$}.
Since we again obtain a negative eigenvalue, the state is once more interpreted as being entangled, as expected. The negativity and logarithmic negativity in the propagation basis can be computed as before, yielding
\begin{equation}
    \tilde{\mathcal{N}}(\rho^{(a)}) = \sum_{\lambda_i<0} |\lambda_i| = |\tilde{c}_1(z)||\tilde{c}_2(z)|=\frac{\abs{\beta}}{2\sqrt{\alpha ^2 + \beta^2}} \;,
\end{equation}
and 
\begin{equation}
    \tilde{E}_{\tilde{\mathcal{N}}}(\rho^{(a)}) = \log_2 |2\tilde{\mathcal{N}}(\rho^{(a)}) + 1| = \log_2\left(\frac{\abs{\beta}}{\sqrt{\alpha ^2 + \beta^2}} + 1\right) \;.
\end{equation}
As with the analogous concurrence measure, the propagation basis negativity measures are static and set entirely by the physical parameters $\alpha$ and $\beta$; independent of the propagation distance. Note also that in both bases, the concurrence and negativity are related, differing only by a factor of two. This is expected, as alluded to earlier, and as we sharpen further below.

In the interaction basis of the two-neutrino system, the respective expressions for negativity and logarithmic negativity are
\begin{equation}
\mathcal{N}(\rho^{(g)})\left[t\right] = \sqrt{\sin^2(2\theta_V)\sin^2{\left(\frac{\Delta m_{ji}^2t}{4E}\right)}\left(1-\sin^2(2\theta_V)\sin^2{\left(\frac{\Delta m_{ji}^2t}{4E}\right)}\right)} \;,
\end{equation}
and
\begin{equation}
E_{\mathcal{N}}(\rho^{(g)})\left[t\right] = \log_2\left(2 \sqrt{\sin^2(2\theta_V)\sin^2{\left(\frac{\Delta m_{ji}^2t}{4E}\right)}\left(1-\sin^2(2\theta_V)\sin^2{\left(\frac{\Delta m_{ji}^2t}{4E}\right)}\right)}+1\right) \;.
\end{equation}
In the propagation basis, the analogous quantities are
\begin{equation}
    \tilde{\mathcal{N}}(\rho^{(g)}) = \frac{\sin \left( 2\theta_V\right)}{2} \;,
\end{equation}
and
\begin{equation}
    \tilde{E}_{\tilde{\mathcal{N}}}(\rho^{(g)}) = \log_2{\left(\sin \left( 2\theta_V\right)+1\right)} \;.
\end{equation}
Just like the concurrence, in the propagation basis, we see that the negativity and logarithmic negativity thresholds in the same basis are set at $\theta_V = \pi/4$ and it is clear that, unlike the axion--photon case, the neutrino oscillation measures can be characterised by simply fixing one external physical parameter, namely, the flavour mixing angle $\theta_V$.

At this juncture, we reiterate our earlier discussion on the speciality of pure bipartite states of Schmidt rank two, as in the current simplified systems of interest, and re-emphasise that a simple structural relation exists among the measures in this special case, namely
\begin{equation}
 \mathcal{C}(z)=2\mathcal{N}\!\qty(\rho^{(a)})(z),\qquad
\SL^{(\gamma;a)}\!\qty(\rho^{(a)})= \mathcal{C}^2(z)=4 \mathcal{N}^2\!\qty(\rho^{(a)})(z) \;.
\label{eq:pure-state-relations}
\end{equation}
Thus, for the analysis of the pure photon state under consideration, concurrence, negativity, and linear entropy are not independent pieces of information---they are, in fact, different monotonic parametrisations of the same underlying Schmidt spectrum. This means that we can identify the resonance condition ($m^2_{\gamma,\parallel} = m^2_a$), the high energy limit of the photon, and the large magnetic background limit ($\beta\gg\alpha$) as among the universal characteristic scenarios corresponding to maximisation, with respect to $\alpha$ and $\beta$. These simple equivalences will no longer necessarily hold true when one considers more realistic scenarios with three-flavour neutrino oscillations, or axion--photon systems with significant magnetic field inhomogeneities and Faraday rotation effects, leading to possible multipartite entanglement effects. 

We must mention here that we also investigated the coherence\,\cite{Plenio:2005cwa} measures (specifically, the $l_1$-norm of coherence and the relative entropy of coherence) for the axion--photon system. We do not present them in detail here since they coincide numerically with the concurrence and the von Neumann entanglement entropy, respectively, in both the interaction and propagation bases.

\subsection{Quantum Mutual Information and Discord}

Quantum discord\,\cite{Ollivier:2001fdq, Brodutch:2016lvw, Modi:2012baj, Modi:2012baj} is a measure of non-classical correlations in a bipartite system that goes beyond entanglement. While entanglement measures the non-separability of a state, quantum discord measures non-classical correlations that may exist even in separable states. This means even if the system is not entangled (i.e., separable), it can still exhibit quantum discord. Physically, quantum discord tries to quantify potential measurement-induced disturbances that may arise when one subsystem is measured. In classical systems, measuring one part of a system does not disturb the other part, but in quantum systems, measurement can change the state of the system, leading to non-classical correlations. We study discord here with a view towards more experimentally realistic studies in the future that leverage quantum information-theoretic measures in the axion--photon system. 

In the axion--photon system, the reduced density matrices are diagonal, so
\begin{equation}
    S(\rho^{(a)}_\gamma) = S(\rho^{(\gamma)}_a) = -\left(|c_1(z)|^2 \log_2 |c_1(z)|^2 + |c_2(z)|^2 \log_2 |c_2(z)|^2\right) \;,
\end{equation}
and for the pure state $\rho^{(a)}$, the von Neumann entropy $S(\rho^{(a)}) = 0$. Therefore, the quantum mutual information\,\cite{Nielsen:2012yss} is found to be 
\begin{equation}
    \operatorname{I(\rho^{(a)})} = 2\cdot S(\rho_{\gamma}^{(a)}) \; .
\end{equation}
The classical correlation\,\cite{Henderson:2001wrr} for the photon subsystem is given by\footnote{Note that $p_k$ is the probability that the outcome due to the projector $\Pi_k^a$ is obtained.}
\begin{equation}
    J_\gamma(\rho^{(a)}) = S(\rho_{\gamma}^{(a)}) - \min_{\{\prod_{k}^{a}\}} \sum_k p_k S(\rho_{\gamma|k}) \; .
\end{equation}
Here, a measurement on the axion subsystem in basis $ \{|0\rangle_a, |1\rangle_a\} $ projects the photon subsystem into the conditional states $ \{ |0\rangle_\gamma $ or $ |1\rangle_\gamma \} $ which are pure states. Hence, the conditional entropy $S(\rho_{\gamma|k})$\,\cite{CerfAdami1997} vanishes and the minimum is also 0. Therefore, the classical correlation is given by
\begin{equation}
    J_\gamma(\rho^{(a)}) = S(\rho_{\gamma}^{(a)}) \; .
\end{equation}
The quantum discord\,\cite{Ollivier:2001fdq}, given by the difference of the quantum mutual information and the classical correlation, $\operatorname{I(\rho^{(a)})}-J_\gamma(\rho^{(a)})$, tries to capture the amount of non-classical correlations present in a quantum system. For the photon subsystem in the interaction basis, it is given by
\begin{align}
    \mathcal{D}_\gamma(\rho^{(a)})[z]=S(\rho_{\gamma}^{(a)}) = &- \left[\frac{\beta^2}{\alpha^2+\beta^2} \sin^2\left(z \sqrt{\alpha^2+\beta^2} \right) \log_2\left( \frac{\beta^2}{\alpha^2+\beta^2} \sin^2\left(z \sqrt{\alpha^2+\beta^2} \right) \right) \right. \nonumber \\
    &\left. + \left( 1 - \frac{\beta^2}{\alpha^2+\beta^2} \sin^2\left( z \sqrt{\alpha^2+\beta^2}  \right) \right)
    \log_2\left( 1 - \frac{\beta^2}{\alpha^2+\beta^2} \sin^2\left( z \sqrt{\alpha^2+\beta^2} \right) \right)
    \right] \; .
\end{align}
The equality to the entanglement entropy in our case suggests that, for fixed physical parameters $\alpha$ and $\beta$, the maximum value attained by the quantum discord measure will again be given by Eq.\,\eqref{eq:maxent}.

In the propagation basis, the analogous expression for the quantum discord is
\begin{align}
    \mathcal{D}_{\tilde{a}_1}(\rho^{(a)}) =-\frac{1}{2}\left(1-\frac{\alpha}{\sqrt{\alpha^2+\beta^2}}\right)\log_2{\left[\frac{1}{2}\left(1-\frac{\alpha}{\sqrt{\alpha^2+\beta^2}}\right)\right]}-\frac{1}{2}\left(1+\frac{\alpha}{\sqrt{\alpha^2+\beta^2}}\right)\log_2{\left[\frac{1}{2}\left(1+\frac{\alpha}{\sqrt{\alpha^2+\beta^2}}\right)\right]} \; .
\end{align}

The equality of discord and entanglement entropy is special to the pure states considered here and should not be expected to survive in more realistic settings. The axion--photon two-mode state becomes mixed due to decoherence in the presence of noise---for instance, from photon absorption in the medium,  fluctuations in the external magnetic field along the propagation direction, or thermal effects. In these dissipative regimes, the quantum discord is known to be more resilient in identifying non-classical behaviour in a system than quantum entanglement, as pure state entanglement ideas do not necessarily carry forward to mixed state entanglement. Thus, we present the quantum discord in the axion--photon system to clarify a possible robust diagnostic of non-classical correlations in the axion--photon system under realistic experimental conditions\,\cite{Brodutch:2016lvw, Modi:2012baj}.

On calculating the expression for quantum discord in the interaction basis of the two-neutrino system, one gets
\begin{align}
    \mathcal{D}_f(\rho^{(g)})[t]= S(\rho_f^{(g)}) &= -\Bigg[
        \sin^2(2\theta_V)\sin^2{\left(\frac{\Delta m_{ji}^2t}{4E}\right)}\log_2 \left( \sin^2(2\theta_V)\sin^2{\left(\frac{\Delta m_{ji}^2t}{4E}\right)} \right) \nonumber \\
        &\quad + \left( 1 - \sin^2(2\theta_V)\sin^2{\left(\frac{\Delta m_{ji}^2t}{4E}\right)} \right) \log_2 \left( 1 - \sin^2(2\theta_V)\sin^2{\left(\frac{\Delta m_{ji}^2t}{4E}\right)} \right)
    \Bigg] \; ,
\end{align}
and in the propagation basis, this corresponds to
\begin{align}
    \mathcal{D}_1(\rho^{(g)}) &= -\left(\cos^2\theta_V \log_2 \cos^2\theta_V + \sin^2\theta_V \log_2 \sin^2\theta_V\right) \;.
\end{align}
The functional features of this measure in the neutrino system will again follow those of the entanglement entropy\,\cite{Blasone:2007wp, Blasone:2007vw}.

\subsection{Capacity of entanglement}
The capacity of entanglement\,\cite{Yao:2010woi, Schliemann:2011zkg,DeBoer:2018kvc}, which describes the fluctuations in the entanglement entropy, is given by
\begin{equation}\label{eq:CEE}
C_E = \mathrm{Tr}(\rho_{\gamma}^{(a)} (\log_2 \rho_{\gamma}^{(a)})^2) - S(\rho_{\gamma}^{(a)})^2
\; .
\end{equation}
This simplifies in the axion--photon system to
\begin{align}\label{eq:CEE_axionphoton}
C_E(z) &= P_{\gamma \to a}(z)\left(1 - P_{\gamma \to a}(z)\right)\left[\log_2 \left(\frac{P_{\gamma \to a}(z)}{1 - P_{\gamma \to a}(z)}\right)\right]^2 \nonumber \\
 \equiv & \frac{\beta^2}{\alpha^2+\beta^2}
\sin^2\left(z \sqrt{\alpha^2+\beta^2} \right)\left(1 - \frac{\beta^2}{\alpha^2+\beta^2}
\sin^2\left(z \sqrt{\alpha^2+\beta^2} \right)\right)\left[\log_2 \left(\frac{\frac{\beta^2}{\alpha^2+\beta^2}
\sin^2\left(z \sqrt{\alpha^2+\beta^2} \right)}{1 - \frac{\beta^2}{\alpha^2+\beta^2}
\sin^2\left(z \sqrt{\alpha^2+\beta^2} \right)}\right)\right]^2 \; .
\end{align}
The capacity of entanglement is symmetric under $P_{\gamma \to a} \leftrightarrow 1-P_{\gamma \to a}$. This suggests that if the actual $P_{\gamma \to a}$ spans $[0,1]$, $C_E(z)$ would have a mirror symmetry centred around $P_{\gamma \to a}=1/2$, as confirmed by Fig.\,\ref{fig:QI_measures_comparison}. In general, depending on the physical parameters $\alpha$ and $\beta$ that determine the maximum transition probability achieved (i.e. $\beta^2/(\alpha^2+\beta^2)$), $C_E(z)$ has a rich functional form across the oscillation time period.

Unlike concurrence or linear entropy, or the other measures calculated, which quantify the magnitude of entanglement, the capacity of entanglement captures the variance in the entanglement spectrum. It vanishes at $P_{\gamma \to a} = 0$ and $P_{\gamma \to a} = 1$, corresponding to separable states, and also at $P_{\gamma \to a} = 1/2$, where the system is maximally entangled. The vanishing of $C_E$ at maximal entanglement and when the state is separable reflects the fact that the Schmidt spectrum becomes uniform, thus eliminating fluctuations. This provides another conceptual perspective for the distinct double-humped profile when compared to other entanglement measures (Fig.\,\ref{fig:QI_measures_comparison}).

As we shall see in the next section, the capacity of entanglement will play a crucial role in the entanglement quantum speed limits. It is therefore pertinent to investigate what the maximum value attained is, for fixed values of $\alpha$ and $\beta$. One finds
\begin{equation}\label{eq:CEMax}
C_{E}^{\,\max}= \begin{cases}\frac{\beta^2}{\alpha^2+\beta^2}\left(1-\frac{\beta^2}{\alpha^2+\beta^2}\right)\left[\log _2\left(\frac{\beta^2}{\alpha^2}\right)\right]^2, & \frac{\beta^2}{\alpha^2+\beta^2}<p_{\star}\;, \\ 0.914 \ldots, & \frac{\beta^2}{\alpha^2+\beta^2} \geq p_{\star} \; .\end{cases}
\end{equation}
Here, $p_{\star}\simeq 0.083 $ is the value, along with $1-p_*\simeq0.917$ (due to the symmetry of the function under $p \leftrightarrow 1-p$), which globally maximises $C_E(p)=p(1-p)\left[\log _2\left(p/1-p\right)\right]^2$. The global maximum is $C_{E}^{global,\max}(p_{\star})\simeq 0.914$. These theoretical deductions are confirmed by the features of $C_E(z)$ in Fig.\,\ref{fig:QI_measures_comparison}.

In the propagation basis, the capacity of entanglement evaluates to
\begin{equation}
\tilde{C}_E =
\frac{\beta^2}{4(\alpha^2 + \beta^2)}
\left[
\log_2\left(\frac{1 - \frac{\alpha}{\sqrt{\alpha^2+\beta^2}}}
{1 + \frac{\alpha}{\sqrt{\alpha^2+\beta^2}}}\right)
\right]^2.
\end{equation}
We note that in general the constant $\tilde{C}_E$ is different from $C_{E}^{\,\max}$. It is perhaps simpler to follow how the physical parameters constrain the bounds of the capacity of entanglement in the propagation basis; the logarithmic term vanishes in both the resonance $\alpha = 0$ and the large magnetic background limit $\beta \gg \alpha$.  Although these characterise the interesting values of this measure, consistent with the behaviour of the other entanglement measures discussed before, we find that the spectrum of values for the capacity of entanglement is complementary to other entanglement measures. This, coupled with the fact that in the interaction basis it has an interpretation in terms of a minimum time required by a bipartite system to generate a certain amount of entanglement\,\cite{Shrimali:2022bvt} means that it is a measure that gives a valuable complementary perspective on entanglement. 

For the two-neutrino system, the corresponding expressions for the capacity of entanglement in the interaction and propagation basis are given by
\begin{equation}
C_E(t)=\sin^2(2\theta_V)\sin^2{\left(\frac{\Delta m_{ji}^2t}{4E}\right)}\left(1 - \sin^2(2\theta_V)\sin^2{\left(\frac{\Delta m_{ji}^2t}{4E}\right)}\right)\left[\log_2\left(\frac{\sin^2(2\theta_V)\sin^2{\left(\frac{\Delta m_{ji}^2t}{4E}\right)}}{1 - \sin^2(2\theta_V)\sin^2{\left(\frac{\Delta m_{ji}^2t}{4E}\right)}}\right)\right]^2.
\end{equation}

\begin{equation}
\tilde{C}_E=\sin^2(2\theta_V)\left[\log_2\left(\cot\theta_V\right)\right]^2.
\end{equation}
Comparing with Eq.\,\eqref{eq:CEMax}, one notes that the maximum value achieved by $C_E(t)$ in the neutrino case is $ \sin^2 (4\theta_V )\log_2^2 (\tan (2\theta_V))$ for $\theta_V < 1/2\,\sin^{-1}\sqrt{p_*} $, and $0.914$ for larger $\theta_V$. We note that for $\tilde{C}_E$, the logarithmic term causes the capacity of entanglement to vanish at $\theta_V = \pi/4$, despite it maximising the standard $\sin^2(2\theta_V)$ prefactor that appears in many entanglement measures for the neutrino system.

Before we close this section on entanglement measures, we comment briefly on the experimental viability of the quantum information-theoretic analysis developed thus far. A growing class of quantum-enhanced axion search experiments\,\cite{Dixit:2020ymh, Yang:2022uil, Yu:2023wsn, Braggio:2024xed, Devlin:2026czx} motivates the single-excitation sector as a relevant regime for axion--photon conversion searches. Furthermore, we note that every measure computed in this section is a function of the conversion probability $P_{\gamma\to a}(z)$, and since the transition probability is (in principle) inferable from the signal statistics of any experiment operating in the single-excitation sector\footnote{Strictly speaking, in single photon detection, one actually uses $P_{a\to \gamma}$, but as pointed out earlier, the analysis and calculations are equivalent.}, the quantum information measures computed here are, at least in principle, experimentally accessible quantities. That being said, realistic strategies for implementing the full quantum information-theoretic analysis would require more than just signal statistics---it could also entail, for instance, careful quantum state tomography. Additionally, the most general analysis may, in principle, also include multi-mode and multi-excitation sectors, with added complexities from magnetic field inhomogeneities, Faraday rotation effects that mix the transverse photon field modes, and so on. Thus, our modest goal here is to provide a simplified quantum information-theoretic framework as a first step towards interpreting the quantum resources that may be exploited in axion--photon experiments.

\section{Quantum speed limits in axion--photon and neutrino oscillations}
\label{sec:axionphotonQSLs}
The quantum speed limits (QSLs)\,\cite{Mandelstam1991, Margolus:1997ih, LevitinToffoli2009} give characteristic time limits for a quantum state to evolve into another, often an orthogonal state. We seek to look at the axion--photon and neutrino systems from the perspective of these different quantum speed limits.

\subsection{Mandelstam--Tamm and Margolus--Levitin quantum speed limits}

The Mandelstam--Tamm bound\,\cite{Mandelstam1991} for the quantum speed limit is closely related to the energy-time uncertainty relation, and gives a characteristic minimum time for unitary time evolution between orthogonal states (see Appendix~\ref{Appendix:QSL}) in terms of the variance of the Hamiltonian as
\begin{equation} \label{eq:qsl_MT}
T^{\perp}_{\text{\tiny{MT}}} = \frac{\pi \hbar}{2\Delta H} \;.
\end{equation}
This assumes that the Hamiltonian is time independent. The alternative formulation by Margolus and Levitin\,\cite{Margolus:1997ih} is in terms of the mean energy (see Appendix~\ref{Appendix:QSL} for more details) above the ground state energy ($E_0$)
\begin{equation} \label{eq:qsl_ML}
T^{\perp}_{\text{\tiny{ML}}} = \frac{\pi \hbar}{2 \left[ \langle H \rangle- E_0 \right]} \; .
\end{equation}

It was subsequently shown that for general $\left(\langle H \rangle- E_0 \right)$  and $\Delta H$ the above combined limits are tight\,\cite{LevitinToffoli2009}. The evolution time between orthogonal states, the so-called orthogonalisation time, for a time-independent Hamiltonian satisfies 
\begin{equation}\label{eq:LTBound}
T \geq T^{\perp}_{\text{\tiny{QSL}}}=\max \left(T^{\perp}_{\text{\tiny{MT}}}, \, T^{\perp}_{\text{\tiny{ML}}} \right) \; .
\end{equation}
The characteristic time defined above specifies the minimum evolution timescale accessible to a quantum system, thereby imposing upper bounds on the rates at which quantum information can be transmitted and processed, as well as on the rate of quantum entropy production (see, for instance, discussions in\,\cite{Deffner:2017cxz} and references therein).

In the case of neutrinos, since the maximum conversion probability $P_{\nu_{f} \rightarrow \nu_{g}}^{\,\max}= \sin^2(2\theta_V) $, true orthogonalisation is only ever achieved, independent of the value of the mass-squared difference $\Delta m_{ji}^2$, for maximal mixing ($\theta^{*}_V=\pi/4$) and the above QSL times coincide
\begin{equation}\label{eq:MTNeu}
T^{\perp}_{\text{\tiny{MT}}}(\nu_f \rightarrow \nu_g) \equiv \frac{\pi \hbar}{\left|\frac{\Delta m_{ji}^2}{ E} \sin \theta^{*}_V \cos \theta^{*}_V \right|} ~~=~~T^{\perp}_{\text{\tiny{ML}}}(\nu_f \rightarrow \nu_g) \equiv \frac{\pi \hbar}{\left|\frac{\Delta m_{ji}^2}{E} \sin^2
\theta^{*}_V \right|} \; .
\end{equation}
This is also trivially equal to the oscillation half-period as dictated by $P_{\nu_{f} \rightarrow \nu_{g}}(t)$, thereby saturating the aforementioned QSL bounds. This is consistent with the general fact that simultaneous saturation of the Mandelstam--Tamm and Margolus--Levitin bounds occurs precisely for two-level superpositions with equal-weight components\,\cite{LevitinToffoli2009, Deffner:2017cxz}, in the present context, of the mass bases $\ket{\nu_i}$ and $\ket{\nu_j}$. In the above context, it is intriguing that in the atmospheric oscillation sector, where $\nu_e$ does not participate significantly and the oscillations are effectively just between the two flavours $\nu_\mu$ and $\nu_\tau$, the atmospheric mixing angle $\theta_{23}$ is very close to the maximal value. A similar observation was also made in the previous section from the point of view of maximum entanglement in the mass basis.

For the axion--photon system, the maximum conversion probability is $P^{\,\max}_{\gamma \rightarrow a}=\beta^2/(\alpha^2+\beta^2)$. Therefore, full orthogonalisation may be achieved independent of the values of $g_{a\gamma\gamma}$ and $B_{\text{ext}}$, but is now dependent on the axion--photon mass-squared difference ($\alpha \propto \Delta m_{\gamma a}^2 $). For full orthogonalisation, the latter manifests as the resonance condition
\begin{equation}
\alpha^*=0 \implies m_{\gamma,\parallel}^2 = m_a^2 \; .
\end{equation} 
Again, both the QSL times coincide in this limit as expected
\begin{equation} \label{eq:photonaxion_qsl_MT}
T^{\perp}_{\text{\tiny{MT}}}(\gamma \rightarrow a) \equiv \frac{\pi}{2\,\abs{\beta}} ~~=~~
T^{\perp}_{\text{\tiny{ML}}}(\gamma \rightarrow a) \equiv \frac{\pi}{2 \left[\alpha^*+\sqrt{\alpha^{*\,2}+\beta^2} \right]}\;.
\end{equation}

It is important to point out an interesting aspect here. Note from Eq.\,\eqref{eq:MTNeu} that the neutrino QSL times vanish in the $\hbar \rightarrow 0$ limit. This is a manifestation of the fact that neutrino oscillations do not persist as a classical phenomenon, as we discussed in Sec.\;\ref{sec:axionphoton}. This is distinct from the axion--photon case above, where intriguingly the QSL characteristic time is actually independent of $\hbar$. This is again a manifestation of the fact that in the axion--photon case the oscillation persists when one takes the classical limit, with high-occupancy numbers and bosonic coherent states. Thus, for the axion--photon mixing, the presence of a minimum characteristic time for distinguishing the two states persists even classically and is more akin to a minimum time for the interconversion of two distinct classical modes. This is of course reflected in the fact that full orthogonalisation, when $\alpha=0$, occurs when in Eq.\,\eqref{eq:osc_prob} we have $\beta\, z_\perp=\pi/2$, which is just the Mandelstam--Tamm characteristic time. 

Let us now consider evolution to non-orthogonal states. This is, in fact, the most general scenario, wherein the neutrino vacuum mixing angle is not necessarily maximal or in the axion--photon system the resonance condition is not necessarily satisfied.

In this case, we have the general Mandelstam--Tamm QSL bounds (see Appendix~\ref{Appendix:QSL}) for the neutrino and axion--photon systems as
\begin{equation}
T \ge T_{\text{\tiny{MT}}}(\nu_f \rightarrow \nu_g) \equiv \frac{\hbar\, \cos^{-1}\left[\sqrt{1-P_{\nu_f \rightarrow \nu_g}(T)}\,\right]}{\left|\frac{\Delta m^2_{ji}}{4E} \sin 2\theta_V\right|} \; ,
\end{equation}
\begin{equation}
    T \ge T_{\text{\tiny{MT}}} (\gamma \rightarrow a) \equiv \frac{\,\cos^{-1}\left[\sqrt{1 - P_{\gamma \rightarrow a}(T)}\,\right]}{\abs{\beta}} \; .
\end{equation}
Consider now the evolution time as that at which the conversion probabilities are at their maximum for the neutrino and axion--photon systems---i.e., the oscillation half-periods ($T^{\,\max}_{\nu}$ and $T^{\,\max}_{a}$). The states are the most distinct at this juncture, and it is therefore a relevant instant for comparison to the QSL characteristic times. Also, it is worth re-emphasising that the respective states at $T^{\,\max}_{\nu}$ and $T^{\,\max}_{a}$ are not fully orthogonal to the respective initial states in general. The corresponding Mandelstam--Tamm QSL times ($T^{\,\max}_{\nu\,,\,\text{\tiny{MT}}}$ and $T^{\,\max}_{a\, , \,\text{\tiny{MT}}}$) then come out to be
\begin{align}
T^{\,\max}_{\nu} &\equiv 
\frac{2 \pi \hbar E } {\Delta m_{ji}^2} \longrightarrow T^{\,\max}_{\nu\,,\,\text{\tiny{MT}}} \equiv \frac{8 \hbar E \theta_V}{\Delta m_{ji}^2 \sin 2\theta_V} \implies \frac{T^{\,\max}_{\nu\,,\,\text{\tiny{MT}}}}{T^{\,\max}_\nu } \in (2 / \pi, 1]~,~\forall\, \Delta m_{ji}^2\,,\,\theta_V \in(0, \pi / 4] \;,\\
T^{\,\max}_a &\equiv \frac{\pi}{2 \sqrt{\alpha^2+\beta^2}}
 \longrightarrow T^{\,\max}_{a\, , \,\text{\tiny{MT}}}\equiv \frac{ \cos^{-1}\left[ |\alpha|/\sqrt{ \alpha^2+\beta^2} \right] }{\abs{\beta}} \implies \frac{T^{\,\max}_{a \,,\,\text{\tiny{MT}}}}{T^{\,\max}_a } \leq~1~,~\forall\, \alpha,\beta \;.
\end{align}
Thus, though there exists generalised Mandelstam--Tamm QSL characteristic times that are finite and well-defined, they are never saturated or attained by the actual oscillation half-periods in both systems. The oscillation half-periods trivially satisfy the generalised Mandelstam--Tamm QSL bounds, for all values of the physical parameters.

Coming to the generalised Margolus--Levitin QSL bounds (see Appendix~\ref{Appendix:QSL} for more details), we have for the neutrino system at the oscillation half-period (where the states are most distinguishable), where the conversion probability is the maximum, the generalised Margolus--Levitin QSL characteristic time as
\begin{equation}
T^{\,\max}_{\nu\,,\,\text{\tiny{ML}}} \equiv \frac{\pi \hbar\left(1-\operatorname{Re} \tilde{A}(T^{\,\max}_{\nu})+\frac{2}{\pi} \operatorname{Im} \tilde{A}(T^{\,\max}_{\nu})\right)}{2\left(\langle H\rangle-E_0\right)}\; .
\end{equation}
At the oscillation half-period, 
\begin{align}
\operatorname{Re} \tilde{A}(T^{\,\max}_{\nu})&=\cos ^2 \theta_V +\sin ^2 \theta_V \cos (\Delta m^2_{ji} T^{\,\max}_{\nu} / 2\hbar E)= \cos 2 \theta_V \; , \nonumber \\
\operatorname{Im} \tilde{A}(T^{\,\max}_{\nu})&= -\sin ^2 \theta_V \sin (\Delta m^2_{ji} T^{\,\max}_{\nu} / 2\hbar E)=0 \;,
\end{align}
which implies that
\begin{equation}\label{eq:GMLNeu}
T^{\,\max}_{\nu\,,\,\text{\tiny{ML}}}=\frac{\pi \hbar E (1-\cos 2 \theta_V)}{ \sin ^2 \theta_V~ \Delta m^2_{ji}}= T^{\,\max}_{\nu}\; .
\end{equation}

Thus, unlike the generalised Mandelstam--Tamm QSL time that is never saturated, the neutrino system does saturate and attain the generalised Margolus--Levitin QSL time during oscillations. The generalised Margolus--Levitin QSL bound is satisfied as an equality for all values of the physical parameters $\Delta m^2_{ji}$ and $\theta_V$.

Considering the combined Levitin-Toffoli type bounds\,\cite{LevitinToffoli2009} of Eq.\,\eqref{eq:LTBound}, note that
\begin{equation}
\frac{T^{\,\max}_{\nu\,,\,\text{\tiny{MT}}}}{T^{\,\max}_{\nu\,,\,\text{\tiny{ML}}}}= \frac{4 \theta_V}{\pi \sin 2 \theta_V}\; ,
\end{equation}
is monotonically increasing for $\theta_V \in(0, \pi / 4]$, and hence the true QSL bound is always set by the generalised Margolus--Levitin characteristic time in Eq.\,\eqref{eq:GMLNeu}.

A similar analysis for the axion--photon gives
\begin{equation}\label{eq:GMLPho}
T^{\,\max}_{a\,,\,\text{\tiny{ML}}}\equiv \frac{\pi}{2 \sqrt{\alpha^2+\beta^2}} = T^{\,\max}_{a} \; ,
\end{equation}
as expected, based on the analogy with the neutrino system. Thus, again, the generalised Margolus--Levitin QSL time is saturated by axion--photon oscillations and is now set by the inverse of the combination of parameters $\sqrt{\alpha^2+\beta^2}$. The combined Levitin-Toffoli bound is also set by the same quantity.

\subsection{Entanglement quantum speed limits}

The concept of quantum speed limits has recently been extended to arbitrary quantum time evolutions\,\cite{Thakuria:2022taf}, for instance, non-unitary evolutions, and also to the growth of observables\,\cite{Mohan:2021rel}. We would like to explore bounds on the rate of entanglement creation in the context of mode entanglement in the axion--photon system.\footnote{During the completion of this work, an application of a similar idea to three-flavour neutrino oscillation experiments appeared in \cite{Jha:2026jef}.} 

\begin{figure}[h]
    \centering
    \includegraphics[width=0.75\textwidth]{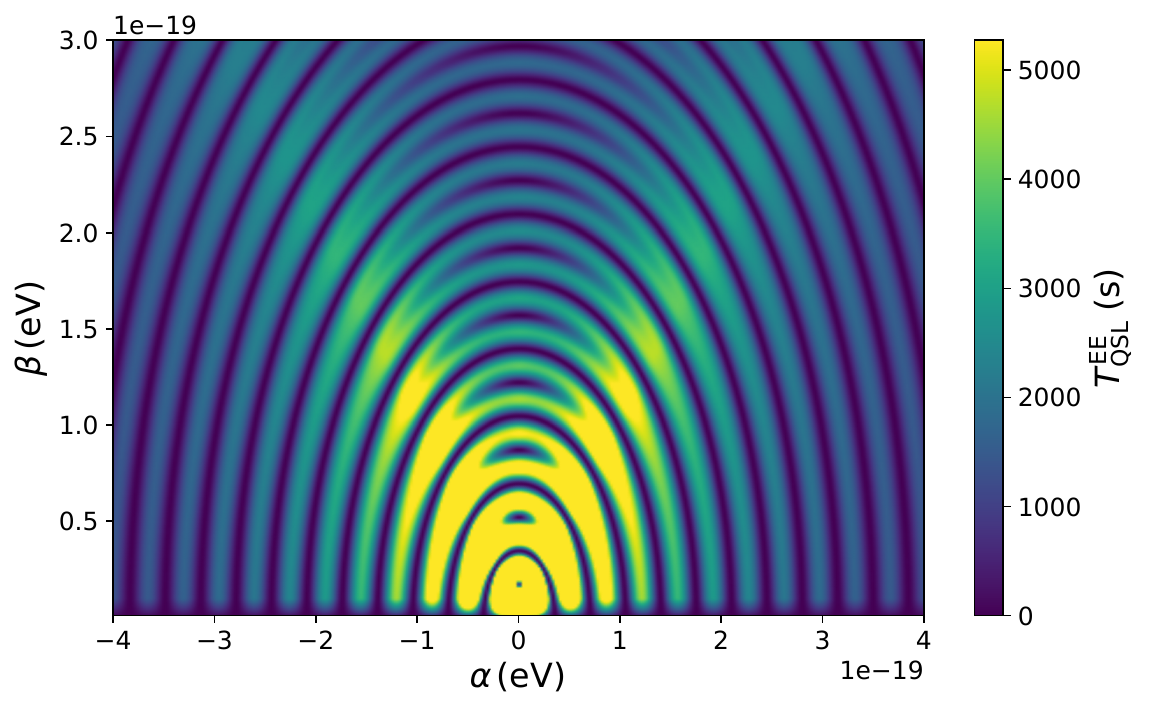}
    \caption{An illustrative figure depicting the entanglement QSL
$T^{\text{\tiny{EE}}}_{\text{\tiny{QSL}}}$
in the $(\alpha,\beta)$ parameter space
for the axion--photon system.}
    \label{fig:entanglementqsl}
\end{figure}


Towards this, we use the notion of an entanglement speed limit \,\cite{Shrimali:2022bvt}. This may be obtained from the generalised Heisenberg-Robertson uncertainty relation \,\cite{Robertson1929} and the Heisenberg equation of motion for the operator $\hat{K}\equiv-\log _2\rho_A$, constructed from the reduced density matrix $\rho_A$. For a bipartite system undergoing unitary dynamics, the corresponding bound on the rate of change of entanglement entropy can be expressed in terms of the capacity of entanglement and the variance of the Hamiltonian. The associated quantum speed limit for entanglement entropy evolution comes out to be \,\cite{Shrimali:2022bvt}
\begin{equation}\label{eq:TEEgeneral}
T^{\text{\tiny{EE}}}_{\text{\tiny{QSL}}} =
\frac{\hbar |S(\rho_A)(T) - S(\rho_A)(0)|}
{2\,\Delta H \, \frac{1}{T}\int_0^T \sqrt{C_E(t)}\,dt} \; ,
\end{equation}
where $C_E(z)$ is the capacity of entanglement as defined in Eq.\,\eqref{eq:CEE_axionphoton}.

For the axion--photon system, using Eqs.\,\eqref{eq:EEap} and\,\eqref{eq:CEE_axionphoton}, we obtain\footnote{Note that due to the ultra-relativistic limit, we can use $z$ and $t$ interchangeably, and so the limits of integration and the integration variable are consistent.}
\begin{equation}\label{eq:TEEaxionphoton}
T^{\text{\tiny{EE}}}_{\text{\tiny{QSL}}}(\gamma \rightarrow a) =
\frac{
\left[-P_{\gamma \rightarrow a}(T) \log_2 P_{\gamma \rightarrow a}(T)- \left( 1 - P_{\gamma \rightarrow a}(T)\right) \log_2\left( 1 - P_{\gamma \rightarrow a}(T)\right)
    \right]}{
2 \, \abs{\beta} \cdot \frac{1}{T}
\int_0^T \sqrt{P_{\gamma \to a}(z)\left(1 - P_{\gamma \to a}(z)\right)\left[\log_2 \left(\frac{P_{\gamma \to a}(z)}{1 - P_{\gamma \to a}(z)}\right)\right]^2}\, dz
} \;.
\end{equation}
From the above equation we note that, roughly speaking, a larger capacity of entanglement, which is related to the fluctuations in the entanglement entropy, implies a shorter characteristic time required to generate a specific amount of entanglement in the system. Also, since both the entanglement entropy and the capacity of entanglement depend on the transition probability $P_{\gamma\to a}(z)$, the dynamics of entanglement are directly controlled by the oscillation frequency $\sqrt{\alpha^2+\beta^2}$. In Fig.\,\ref{fig:entanglementqsl}, we illustrate how the entanglement QSL $T^{\text{\tiny{EE}}}_{\text{\tiny{QSL}}}$ varies as a function of the physical parameters $\alpha$ and $\beta$.

\begin{figure*}[h!]
    \centering
    \includegraphics[width=0.55\textwidth]{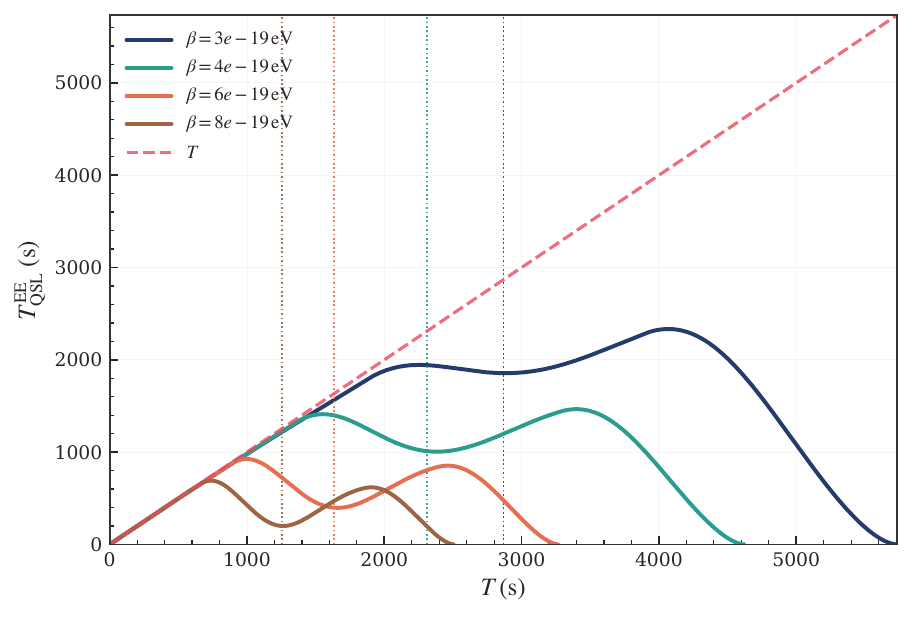}
    \includegraphics[width=0.55\textwidth]{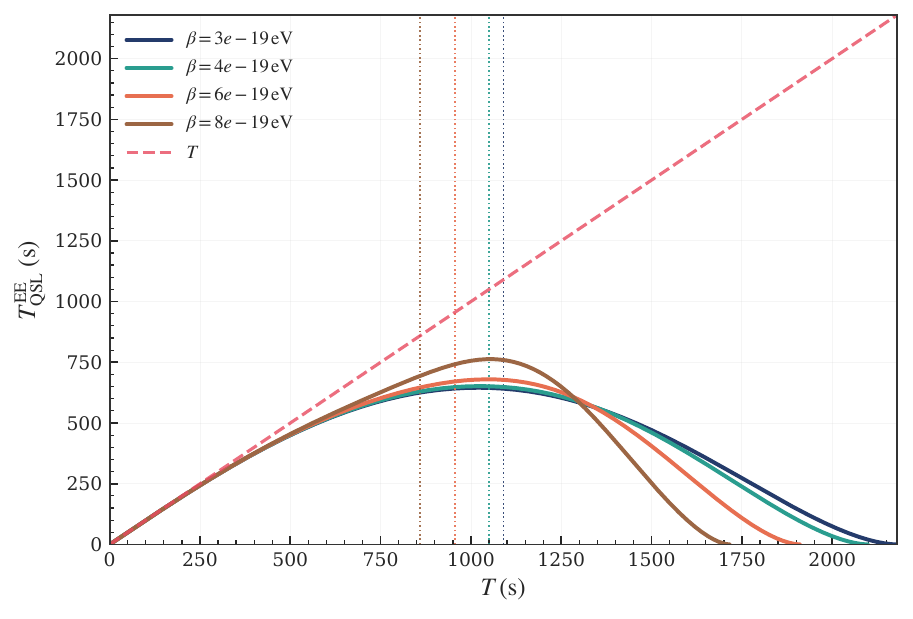}
    \caption{The entanglement QSL time $T^{\text{\tiny{EE}}}_{\text{\tiny{QSL}}}$ plotted as a function of the evolution time $T$, over the respective full oscillation time periods for different values of $\beta$. In the plots $\alpha = 2\times10^{-19}\,\mathrm{eV}$ (top) and $\alpha =9\times10^{-19}\,\mathrm{eV}$ (bottom), corresponding to the regimes $\alpha < \beta$ and $\alpha > \beta$, respectively. The vertical lines correspond to the respective oscillation half-periods, where the transition probabilities are at their maxima. The diagonal, dashed line denotes a reference line $T^{\text{\tiny{EE}}}_{\text{\tiny{QSL}}}=T$. One observes that the entanglement bound is attained and saturated till some $T$, making the bound tight, and beyond that point the bound is weak with $T > T^{\text{\tiny{EE}}}_{\text{\tiny{QSL}}}$. }
    \label{fig:entanglementqslTtight}
\end{figure*}

Whether the entanglement QSL time gives an operationally meaningful characteristic time will depend on how tight the bound is and if the actual time for achieving a specific amount of entanglement saturates the bound. Also of interest is the general behaviour of the characteristic time for entanglement across various domains of the physical parameters $\alpha$ and $\beta$.

In Fig.\,\ref{fig:entanglementqslTtight}, we plot the entanglement QSL times versus the evolution times $T$ over complete oscillation time periods. One intriguing observation is that the bound is saturated up to a point and then becomes weak. We consider two regimes in the subplots, $\alpha < \beta$ (top plot) and $\alpha > \beta$ (bottom plot). The functional forms reflect distinct global characteristics.

In the $\alpha < \beta$ regime (Fig.\,\ref{fig:entanglementqslTtight}, top plot), where $\beta$ dominates, it is seen that if one decreases the mixing parameter $\beta$, the period for which the bound is tight increases. This means that in this regime, the entanglement QSL time is significantly governed by the mixing strength $\beta \sim g_{a\gamma\gamma} B_{\text{ext}}$, and weaker axion couplings $g_{a\gamma\gamma}$ lead to efficient entanglement generation over a longer time period. 

In addition, we see that the contribution from the integral of $\sqrt{C_E(z)}$ initially increases steeply and then approximately flattens out with some mild modulation suggestive of the behaviour seen in Fig.\,\ref{fig:QI_measures_comparison}. The double-humped functional behaviour observed in Fig.\,\ref{fig:entanglementqslTtight} (top plot) is primarily from the $S(\rho_A)(T)$ term in the numerator. For the $\alpha$ and $\beta$ values in this regime, the maximum transition probability attained at oscillation half-period satisfies $\beta^2/(\alpha^2+\beta^2) > 1/2$, and the $S(\rho_A)(T)$ profile at this oscillation half-period is already close to the full profile spanning the $[0,1]$ range in Fig.\,\ref{fig:QI_measures_comparison}. This gives the first hump-like feature before the oscillation half-period, and then, by periodicity, a similar feature after. The behaviour in this regime is also clear from Eq.\,\eqref{eq:maxent}. Of course, in addition, this feature is modulated by the $\sqrt{C_E(z)}$ integral, thereby slightly deforming the features and making it not exactly mirror-symmetric about the oscillation half-period.

In contrast, in the $\alpha > \beta$ regime (Fig.\,\ref{fig:entanglementqslTtight}, bottom plot), the features are dominated by $\alpha$, and consequently in a way by the axion mass-squared value; for instance, assuming fixed experimental parameters that control the effective photon mass. In this regime, the saturation of the entanglement QSL bound is found to be almost independent of the $\beta$ values. This means that the bound is tight for the same duration irrespective of the axion coupling $g_{a\gamma\gamma}$, for a fixed external field. It is also noted here that the bound is tight and saturated only for a short duration, compared to the $\alpha < \beta$ case. The single-humped feature is again a manifestation of the $S(\rho_A)(T)$ term in the numerator modulated by the integral of $\sqrt{C_E(z)}$ in the denominator. In this regime, the maximum transition probability, achieved at the oscillation half-period is $\beta^2/(\alpha^2+\beta^2) < 1/2$. Therefore, the $S(\rho_A)(T)$ profile over the full oscillation time period approaches a maximum value as given in Eq.\,\eqref{eq:maxent} and then decreases, giving a single hump-like feature. The feature is again softened by the integral of $\sqrt{C_E(z)}$ in the denominator, making it slightly asymmetric about the oscillation half-period.

\section{Summary and conclusions}\label{sec:summaryconclusions}

In this work, we initiated a study of axion--photon oscillations in the quantum, low-occupancy regime from a quantum information-theoretic viewpoint. In the presence of an external magnetic field, the axion--photon system in the single-excitation sector manifests mode entanglement, rather than particle entanglement, in direct analogy with single-photon mode entanglement in quantum optics, and with previous treatments of neutrino oscillations as a mode entangled system\,\cite{Blasone:2007wp,Blasone:2007vw}. In this domain, one may investigate various quantum information measures, which may be relevant to new state-of-the-art experimental methods leveraging quantum measurements and quantum sensing\,\cite{Degen:2017QuantumSensing,Dixit:2020ymh,Yang:2022uil,Yu:2023wsn,Braggio:2024xed,Zheng:2024kxn,Devlin:2026czx}.

We investigated in detail a class of quantum information measures in the axion--photon system, and characteristic features related to their evolution in the interaction basis along with the thresholds for their extremisation. Wherever new and unaddressed in the existing literature, we also discussed corresponding aspects in the two-flavour neutrino oscillation context. In the interaction basis, which is the basis directly tied to photon and axion detection, many of the standard pure-state entanglement measures are controlled just by the conversion probability $P_{\gamma\to a}$. In particular, the entanglement entropy reduces to the binary entropy of this probability, while the concurrence, linear entropy and negativity are different monotonic parametrisations of the same Schmidt spectrum, giving complementary perspectives. Consequently, their extrema are attained in the same physical regimes. Maximal entanglement is possible when the dynamics can generate an equal-weight axion--photon superposition. For instance, maximisation may occur with sufficiently strong mixing relative to detuning; in particular, the resonance condition $m_{\gamma,\parallel}^2=m_a^2$ and the large-mixing regime $\beta\gg \alpha$ are among the characteristic regimes where this may occur. These are also precisely the regimes in which the axion--photon conversion probability is enhanced, thereby tying the quantum information structure directly to the experimentally relevant parameter space of axion searches. For fixed physical parameters $\alpha$ and $\beta$, the extrema of the measures were found to occur at characteristic threshold values.

We also evaluated quantum discord, which quantifies non-classical correlations stemming from measurement-induced disturbances. Although it identically reduces to the entanglement entropy for the pure bipartite states considered here, establishing this measure might be essential for future experimental analyses, where discord is expected to be a more robust diagnostic of non-classical behaviour against realistic environmental decoherence. Furthermore, beyond the magnitude-based monotones, we also consider the capacity of entanglement\,\cite{Yao:2010woi, Schliemann:2011zkg,DeBoer:2018kvc} which provides an especially useful complementary diagnostic in the axion--photon system---unlike the entanglement measures, it quantifies the variance of the entanglement spectrum, rather than just the magnitude of entanglement itself. It displays a rich functional structure in the axion--photon system. Additionally, it also plays a role in the entanglement quantum speed limit.

We also studied the Mandelstam--Tamm and Margolus--Levitin quantum speed limits\,\cite{Mandelstam1991, Margolus:1997ih, LevitinToffoli2009} in the axion--photon and neutrino systems. For neutrinos, orthogonalisation is possible only at maximal mixing, in which case the Mandelstam--Tamm and Margolus--Levitin times coincide with the oscillation half-period. For axion--photon oscillations, full orthogonalisation occurs only at resonance, and the corresponding limiting time is set by the inverse mixing strength, $1/\beta$. The characteristic quantum speed limit time is found to be independent of $\hbar$ in the axion--photon case. The fact that this characteristic time is independent of $\hbar$ is not a paradox; rather, it reflects the persistence of axion--photon oscillations in the classical wave limit. It is therefore better understood as the minimum interconversion time between two coupled bosonic modes, with the quantum single-excitation interpretation furnishing the corresponding mode entanglement description. For non-orthogonal evolution, relevant away from maximal neutrino mixing or away from the axion--photon resonance condition, it is found that the generalised Mandelstam--Tamm bounds remain finite but are not saturated by the oscillation half-periods. In contrast, the generalised Margolus--Levitin bounds are saturated at the points of maximal conversion in both systems. Thus, for the oscillation maxima considered here, the Margolus--Levitin-type bound is found to give the operative tight quantum speed limit, while the Mandelstam--Tamm bound is generally weaker except in the fully orthogonalising limits. For axion--photon oscillations, the resulting characteristic timescale is controlled by $\sqrt{\alpha^2+\beta^2}$, making explicit how the interplay between detuning, axion mass, photon effective mass, magnetic field strength and axion--photon coupling enters the speed-limit structure. 

Finally, we studied an entanglement quantum speed limit\,\cite{Shrimali:2022bvt} for the axion--photon system, governed by the entanglement entropy, the Hamiltonian variance and the time-averaged square root of the capacity of entanglement. This bound reveals a non-trivial dependence on the physical mixing parameters. In the $\alpha<\beta$ regime, the behaviour is primarily governed by the mixing strength, and the entanglement-speed profile inherits the double-humped structure associated with the entropy and capacity of entanglement. In the $\alpha>\beta$ regime, the detuning dominates, the profile becomes effectively single-humped, and the duration over which the bound is tight is comparatively shorter and almost insensitive to $\beta$. These results show that the quantum information structure of axion--photon oscillations is not merely a formal rewriting of the transition probability, but encodes a physically interpretable hierarchy of regimes tied to resonance, detuning and magnetic-field-induced mixing.

The analyses presented here broach the simplest framework for quantum information measures in axion--photon oscillations---a single-mode, single-excitation, two-state system in a fixed external magnetic background. Within this simplified setting, every measure is ultimately expressible in terms of experimentally meaningful conversion probabilities and system parameters. This makes the framework naturally aligned with the emerging program of quantum-enhanced and single-photon axion searches\, \cite{Dixit:2020ymh,Yang:2022uil,Devlin:2026czx}. Extensions to multi-mode sectors, multi-excitation states, realistic noise, absorption, magnetic-field inhomogeneities, and Faraday-mixing effects would be important next steps. Viewed from this standpoint, the present work contributes to a simplified baseline formulation and study of the axion--photon system, from a quantum information perspective, and furnishes a bridge between axion phenomenology and operational quantum information theory.

\section{Acknowledgements}
AD acknowledges fruitful discussions with Simon White and Lucas Ostrowski. PT acknowledges support from the IISER Pune Summer Student Programme. AT would like to thank Amol Dighe for discussions. 

\appendix
\section{Axion--photon oscillations and evolution}\label{Appendix:AxionPhotonEquations}

\subsection{Linearising the equation}

Eq.\,\eqref{eq:propeq} may be linearised using a standard technique\,\cite{PhysRevD.37.1237}, by separating the common faster oscillations from the distinct slow variations of the amplitudes. It is essential to note the conditions required for the linearisation, as they also are related to why we pick the single-mode single-excitation sector for the quantum regime.

First, the wave ansatz $a(z,t) = a(z)e^{-i\omega t}$ and $A_{\parallel}(z,t) = A_{\parallel}(z)e^{-i\omega t}$ has already separated the spatial amplitudes from the oscillations in time, and also fixed the frequencies (and consequently, the wave number) of the propagating fields. In the ultra-relativistic limit ($k \approx \omega$), one may then approximate
\begin{equation}
    \omega^2 + \partial_z^2 = (\omega + i\partial_z)(\omega - i\partial_z)\simeq 2\omega \, (\omega + i\partial_z) \; .
\end{equation}
Then the equation approximates to the form
\begin{equation}
 i\partial_z \mathbb{\Psi}(z) = \left( \frac{\mathbb{M}^2}{2\omega} - \omega \right) \cdot \mathbb{\Psi}(z) \;.
\end{equation}
With a field redefinition $\mathbb{\Psi}(z) \rightarrow e^{i\omega z} \mathbb{\Psi}(z)$, one may further eliminate the term proportional to $\omega$ and obtain for the oscillation equation of the effective two-level system
\begin{equation}\label{eq:schrodingereq-app}
i \partial_z \begin{pmatrix} A_\parallel(z) \\ a(z) \end{pmatrix} = \frac{1}{2\omega}
    \begin{pmatrix}
        m_{\gamma,\parallel}^2 & g_{a\gamma\gamma} \omega B_{\text{ext}} \\
        g_{a\gamma\gamma} \omega B_{\text{ext}} & m_a^2
    \end{pmatrix} \cdot \begin{pmatrix} A_\parallel(z) \\ a(z) \end{pmatrix} \;.
\end{equation}
As discussed in the main text, we note that the basis in Eq.\,\eqref{eq:schrodingereq-app} is the interaction basis (what would be referred to as the flavour basis in the context of neutrino oscillations). In contrast, the basis that would diagonalise the above effective Hamiltonian will be referred to as the propagation basis (analogue of the mass basis for neutrinos).

\subsection{Effective Hamiltonian operator for the single-excitation sector}

The Hamiltonian operator in the quantum regime, as denoted in Eq.~\eqref{eq:effhamiltonian}, has the same form as the classical mixing matrix in Eq.\,\eqref{eq:hamiltonian}. We follow the quantum field theoretic treatment of the axion--photon system (more generally, see discussions in\,\cite{Ikeda:2025qac, Binger:1999nj, Blasone:2001du,Capolupo:2019xyd}) to motivate this.

Following a standard treatment\,\cite{Ikeda:2025qac,Capolupo:2019xyd} of the axion interaction term $\mathcal{L}\supset a F^{\mu\nu} \tilde{F}_{\mu\nu}$, one may obtain the relevant interaction part of the Hamiltonian operator. For a single mode $\vec{k}$ (fixing the mode is equivalent and consistent with fixing the frequency in the classical wave ansatz) in the ultra-relativistic limit ($\omega_{a} = \omega_{\gamma} \approx k$), it is of the form 
\begin{equation}
    \hat{H} \supset \frac{-ig_{a\gamma\gamma}B_{\text{ext}}}{2}\left(\hat{b}^{\dagger}\hat{c} - \hat{c}^{\dagger}\hat{b}\right) \; ,
\end{equation}
rewritten in terms of the oscillator conventions in Eq.~\eqref{eq:effhamiltonian}. To find the full effective Hamiltonian for the single-mode, single-excitation sector, we just need to complement this with the free particle Hamiltonian for each of the axionic and photonic fields, which is
\begin{equation}
    H_{\text{\tiny{free}}} = \omega_{\gamma}\;\hat{b}^{\dagger} \hat{b} + \omega_a\;\hat{c}^{\dagger} \hat{c} \;.
\end{equation}
To bring it into a more suggestive form, without loss of generality, we rescale $ \hat{b} \equiv -i \hat{b}$\;\footnote{Note that this preserves the bosonic oscillator commutation limits, and is simply a statement of changing the basis vector by a phase (here, $-i$).}. This amounts to simply adding a phase to the $|\gamma\rangle$ state that spans this reduced Hilbert space. With this, our total effective Hamiltonian becomes
\begin{equation}
    \hat{H}_{\text{\tiny{eff}}} = \omega_{\gamma}\;\hat{b}^{\dagger} \hat{b}
    + \omega_{a}\;\hat{c}^{\dagger} \hat{c}
    + \frac{g_{a\gamma\gamma}B_{\text{ext}}}{2}\left(\hat{b}^{\dagger} \hat{c} + \hat{c}^{\dagger} \hat{b}\right) \;.
\end{equation}
Lastly, we consider the ultra-relativistic limit which gives us
\begin{equation}
    \omega_{\gamma} \approx \omega + \frac{m^2_{\gamma, ||}}{2\omega},\qquad \omega_{a} \approx \omega + \frac{m^2_a}{2\omega} \;.
\end{equation}
The leading order $\omega$ can be absorbed out by a suitable phase redefinition when we consider the Schr\"odinger equation, as any term proportional to $\omega|\Psi\rangle$ can be taken care of with the time derivative $i\partial_t(e^{i\omega t}|\Psi\rangle)$. Thus, we are left with (again, restoring for clarity all the $\hbar$ factors explicitly)
\begin{equation}
    \hat{H} = \frac{1}{2\omega}\begin{pmatrix}
        m_{\gamma,\parallel}^2/\hbar & g_{a\gamma\gamma} \hbar \omega B_{\text{ext}} \\
        g_{a\gamma\gamma} \hbar \omega B_{\text{ext}} & m_a^2/\hbar
    \end{pmatrix} \; ,
\end{equation}
which has the same form as the mixing matrix present in the classical linearised (Schr\"odinger-like) equations of motion. Note that this relation is for the matrix representation of $\hat{H}$ in the appropriate $|\gamma\rangle, |a\rangle$ basis, while $\mathbb{H}$ is the classical mixing matrix for the fields $A_{\parallel}(z)$ and $a(z)$. Observe that when we write the Schr\"odinger equation for this system, the $\hbar$ drops leading to the same equation as in the classical regime, thus implying that the oscillations persist in the classical limit as well. This is, in fact, the leading-order transition probability that is shown in Ref.\,\cite{Ikeda:2025qac}, and is consistent with our effective treatment in this paper.

\subsection{Evolution of product of coherent states}

Consider that a system is governed by a Hermitian, total-number-conserving Hamiltonian containing at most mixed quadratic terms of creation and annihilation operators of the particle species present, say,
\begin{equation}
    \hat{H}=\sum_{i j} h_{i j} \hat{a}_i^{\dagger} \hat{a}_j, \quad h_{i j}=h_{j i}^* \; .
\end{equation}
Let the initial state be a direct product of coherent states 
\begin{equation}
    |\vec{\alpha}\rangle=\bigotimes_i\left|\alpha_i\right\rangle ~~;~~~~~\hat{a}_i|\vec{\alpha}\rangle=\alpha_i|\vec{\alpha}\rangle \; .
\end{equation}
From the Heisenberg equations of motion, we get
\begin{equation}
    \hat{U}^{\dagger}(t)\, \hat{a}_i \, \hat{U}(t)=\sum_j M_{ij}(t) \, \hat{a}_j \; ,
\end{equation}
where we have defined the time-evolution operator $\hat{U}(t)=e^{-i \hat{H} t}$ and $M(t)=e^{-i h t}$. Using the above result one then obtains
\begin{equation}
\hat{a}_i \, \hat{U}(t)|\vec{\alpha}\rangle=\hat{U}(t)\,\left(\hat{U}^{\dagger}(t)\, \hat{a}_i \, \hat{U}(t)\right)|\vec{\alpha}\rangle = \hat{U}(t) \sum_j M_{ij} \, \hat{a}_j|\vec{\alpha}\rangle=\left(\sum_j M_{ij}  \, \alpha_j \right) \hat{U}(t)|\vec{\alpha}\rangle \; .
\end{equation}
Hence, the time evolved state $\hat{U}(t)|\vec{\alpha}\rangle$, starting from an initial state $|\vec{\alpha}\rangle$ that is a product of coherent states, is a simultaneous eigenstate of all the annihilation operators $\hat{a}_i$, with eigenvalues $\sum_j M_{ij}  \, \alpha_j$. 

Thus, an initial product of coherent states evolves into a product of coherent states when the Hamiltonian conserves the total particle number and contains at most quadratic terms, possibly mixed, of the creation and annihilation Fock state operators. The Hamiltonian in Eq.\,\eqref{eq:effhamiltonian} is precisely of this form and if one were to start with a coherent state for the photon, in the large-occupancy classical limit, it will only evolve into a product of coherent states of the photon and axion, which is classical, and never generating any quantum entanglement.

\section{Quantum speed limit}\label{Appendix:QSL}
\subsection{The Fubini--Study metric for time evolution}

We begin by defining the infinitesimal distance in Hilbert space with respect to time evolution
\begin{equation}
ds^2 = \| \psi(t+\Delta t) - \psi(t) \|^2 \; .
\end{equation}
Expanding to leading order in $\Delta t$, we get
\begin{align}
ds^2 
&= \| \dot{\psi}(t)\, \Delta t \|^2 = \langle \dot{\psi}(t) \mid \dot{\psi}(t) \rangle \, \Delta t^2 \; .
\end{align}
This quantity is real and positive semidefinite. However, one cannot directly interpret this as a suitable metric on the projective Hilbert space, as it changes, for instance, under gauge transformations. More generally, what we want is the notion of a metric on a complex projective space\,\cite{Fubini1904,Study1905}, invariant under multiplication by phases $e^{i \theta}$.

For simplicity, consider a constant shift of the potential
\begin{equation}
V(x) \to V(x) + V_0 \;.
\end{equation}
Then the wavefunction transforms as
\begin{equation}
\psi(x,t) \to e^{-i \Lambda(t)} \psi(x,t) \; ,
\end{equation}
where we have defined $\Lambda(t) = V_0 t$.

Under this transformation,
\begin{align}
\langle \dot{\psi}(t) \mid \dot{\psi}(t) \rangle 
&\to \left\langle \partial_t \left(e^{-i\Lambda(t)} \psi \right) \,\middle|\, 
\partial_t \left(e^{-i\Lambda(t)} \psi \right) \right\rangle \; .
\end{align}
and we therefore obtain
\begin{equation}\label{eq:p1}
\langle \dot{\psi} \mid \dot{\psi} \rangle 
\to \langle \dot{\psi} \mid \dot{\psi} \rangle 
+ 2 i \dot{\Lambda} \langle \psi \mid \dot{\psi} \rangle 
+ \dot{\Lambda}^2 \; .
\end{equation}
Now, from
\begin{equation}
\partial_t \langle \psi \mid \psi \rangle = 0 \; ,
\end{equation}
we note that
\begin{equation}
i \langle \psi \mid \dot{\psi} \rangle \in \mathbb{R} \; .
\end{equation}
Also, under the transformation being considered, we have
\begin{align}\label{eq:p2}
i \langle \psi \mid \dot{\psi} \rangle
&\to i \langle e^{-i\Lambda} \psi \mid \partial_t (e^{-i\Lambda} \psi) \rangle= i \langle \psi \mid \dot{\psi} \rangle + \dot{\Lambda} \; .
\end{align}
Finally, from Eqs.\,\eqref{eq:p1} and \eqref{eq:p2}, including a factor of $4$, to be consistent with the usual convention in most literature, we see that 
\begin{align}\label{eq:FBSMetric}
ds^2 
&:= 4 \left[\langle \dot{\psi} \mid \dot{\psi} \rangle 
- \left( i \langle \psi \mid \dot{\psi} \rangle \right)^2\right] dt^2 \;,
\end{align}
is invariant under the gauge transformation and is a suitable metric in relation to time evolution. This is the Fubini--Study metric\,\cite{Fubini1904,Study1905} on the space of quantum states (rays).

\subsection{Geodesic distance between time-evolved states}

The expression for the ``shortest" geodesic distance between the states $\psi(0)$ and the time-evolved $\psi(T)$ may be motivated by considering the Fubini--Study metric and the geodesic great circle path between them in the two-dimensional subspace spanned by $\psi(0)$ and the orthogonal state $\psi_{\perp}(0)$; defined implicitly through $\left\langle\psi(0) \mid \psi_{\perp}(0)\right\rangle=0$.

Assume that $\psi(0)$ and $\psi(T)$ are normalised. In the two-dimensional subspace as defined above, up to an irrelevant phase, let
\begin{equation}
    \langle\psi (0) \mid \psi(T) \rangle=\cos \vartheta(T), \quad 0 \leq \vartheta(T) \leq \frac{\pi}{2} \; .
\end{equation}
We may then write, for some suitable $\left|\psi_{\perp}(0)\right\rangle$,
\begin{equation}
|\psi(T)\rangle=\cos \vartheta(T)\,|\psi(0)\rangle+\sin \vartheta(T) \, \left|\psi_{\perp}(0)\right\rangle \; .
\end{equation}

Given this, a natural choice for the interpolating geodesic great circle curve in the two-dimensional subspace is
\begin{equation}
|\zeta(\lambda)\rangle=\cos \lambda \, |\psi(0)\rangle+\sin \lambda \, \left|\psi_{\perp}(0)\right\rangle, \quad 0 \leq \lambda \leq \vartheta(T) \; ,
\end{equation}
with the identifications $|\zeta(0)\rangle=|\psi(0)\rangle$ and $|\zeta(\vartheta)\rangle=|\psi(T)\rangle$. $\lambda$ is the affine parameter tracing the geodesic great circle path. Note that this, in general, will not be the curve traced in the projective Hilbert space by the actual $\psi(t)$, under time evolution, starting from the initial state $\psi(0)$. Or, in other words, they will only coincide if the Hamiltonian is specifically such that the time evolution coincides with the geodesic path in the two-dimensional subspace.

One notes that
\begin{equation}
\frac{d}{d \lambda}|\zeta(\lambda)\rangle=-\sin \lambda \, |\psi(0)\rangle+\cos \lambda \, \left|\psi_{\perp}(0)\right\rangle \; ,
\end{equation}
and 
\begin{equation*}
\left\langle\left.\frac{d \zeta}{d \lambda} \right\rvert\, \frac{d \zeta}{d \lambda}\right\rangle=1~~~,\quad \left\langle\zeta \left\lvert\, \frac{d \zeta}{d \lambda}\right.\right\rangle=0\;.
\end{equation*}
Then, from the Fubini--Study metric in Eq.\,\eqref{eq:FBSMetric} with $\lambda$ as the affine parameter over the geodesic curve, instead of `$t$', the length of the geodesic great circle connecting $\psi(0)$ and $\psi(T)$ is given by  
\begin{equation}
S_0=\int_0^{\vartheta(T)} 2 \, d\lambda =2\, \vartheta(T)\,\equiv\, 2 \cos ^{-1}(|\langle\psi(0) \mid \psi(T)\rangle|) \; .
\end{equation}
This may also be intuitively understood as the ``shortest distance" since it is the angle swept, up to a normalisation, by the ray $|\psi(0)\rangle$ along an arc to reach the ray $|\psi(T)\rangle$. The angle swept, $\vartheta(T)$, is referred to as the Bures angle. 
\subsection{Mandelstam--Tamm bound}
From the Fubini--Study metric, using
\begin{align} 
\langle\dot{\psi}|\dot{\psi}\rangle &= \frac{1}{\hbar^2}\langle\psi|H(t)^2|\psi\rangle = \frac{1}{\hbar^2}\langle H(t)^2\rangle \; ,\\ 
|\langle\psi|\dot{\psi}\rangle|^2 &= \left| -\frac{i}{\hbar}\langle\psi|H(t)|\psi\rangle \right|^2 = \frac{1}{\hbar^2} \langle H(t)\rangle^2 \;,
\end{align}
we have
\begin{equation}
ds^2 = \frac{4}{\hbar^2} \left[ \langle H(t)^2\rangle - \langle H(t)\rangle^2 \right] dt^2 \;.
\end{equation}
In the above expression, $\Delta H^2 = \langle H(t)^2\rangle - \langle H(t)\rangle^2$ is the variance of the Hamiltonian. Comparing this to the geodesic distance between the initial state $\ket{\psi(0)}$ and the time-evolved state $\ket{\psi(T)}$, we have a general bound
\begin{equation}
\frac{2}{\hbar}\int_0^T\Delta H(t)dt \geq   2\cos^{-1}\!\big(|\langle\psi(0)|\psi(T)\rangle|\big)\equiv 2\,\vartheta(T) \;.
\end{equation}

Let us consider the special case where the time-evolved state $\ket{\psi(T)}$ is fully orthogonal to the initial state $\ket{\psi(0)}$. For a time-independent Hamiltonian and evolution to a fully orthogonal state ($\vartheta(T)=\pi/2$), one then obtains the Mandelstam--Tamm bound\,\cite{Mandelstam1991}.
\begin{equation}
T \ge T^{\perp}_{\text{\tiny{MT}}} = \frac{\pi\hbar}{2\Delta H}\;.
\end{equation}
Note that the Mandelstam--Tamm bound on the evolution time is in terms of the variance of the Hamiltonian.

\subsection{Margolus--Levitin bound}

One may alternatively derive a bound on the time for evolving from a state $\ket{\psi(0)}$ to a general state $\ket{\psi(T)}$ in terms of the expectation value of the Hamiltonian\,\cite{Margolus:1997ih}. Consider the state at $t=0$ expanded in terms of the energy eigenbasis $\ket{E_n}$
\begin{equation}
\ket{\psi(0)}=\sum_n c_n \ket{E_n},
\end{equation}
with $\sum_n |c_n|^2=1$. Then,
\begin{equation}
|\psi(t)\rangle=e^{-i H t / \hbar}|\psi(0)\rangle=\sum_n c_n e^{-i E_n t / \hbar}\left|E_n\right\rangle \;.
\end{equation}
Define
\begin{equation}
\epsilon_n = E_n - E_0 \;,
\end{equation}
where $E_0$ is the ground state energy of the system. Note that with this shift we have $\epsilon_n \geq 0,\; \forall n$. Then,
\begin{equation}
A(t) \equiv\langle\psi(0) \mid \psi(t)\rangle=e^{-i E_0 t / \hbar} \sum_n \left|c_n\right|^2 e^{-i \epsilon_n t / \hbar}\equiv e^{-i E_0 t / \hbar} \tilde A(t)\;,
\end{equation}
with,
\begin{equation}
\Re \tilde A(t)=\sum_n |c_n|^2 \cos\!\left(\frac{\epsilon_n t}{\hbar}\right),
\qquad
\Im \tilde A(t)=-\sum_n |c_n|^2 \sin\!\left(\frac{\epsilon_n t}{\hbar}\right).
\end{equation}
Using the inequality
\begin{equation}
\cos x \ge 1-\frac{2}{\pi}(x+\sin x)
\qquad \forall \, x\ge 0\;,
\end{equation}
we obtain
\begin{equation}
\Re \tilde A(t)\ge
1-\frac{2t}{\pi\hbar}\sum_n |c_n|^2\epsilon_n
+\frac{2}{\pi}\Im \tilde A(t) \;.
\end{equation}
Since
\begin{equation}
\sum_n |c_n|^2\epsilon_n=\langle H\rangle-E_0 \;,
\end{equation}
this becomes
\begin{equation}
\Re \tilde A(t)-\frac{2}{\pi}\Im \tilde A(t)
\ge
1-\frac{2\bigl(\langle H\rangle-E_0\bigr)t}{\pi\hbar} \;.
\end{equation}

Considering an evolution from an initial state $\ket{\psi(0)}$ to a general time-evolved state $\ket{\psi(T)}$, not necessarily completely orthogonal to the initial state, in time $T$, the above may be inverted to give
\begin{equation}
T \geq \frac{\pi \hbar\left(1-\operatorname{Re} \tilde{A}(T)+\frac{2}{\pi} \operatorname{Im} \tilde{A}(T)\right)}{2\left(\langle H\rangle-E_0\right)} \; .
\end{equation}
This is the generalised Margolus--Levitin bound, with the identification of a generalised Margolus--Levitin time
\begin{equation}
T_{\text{\tiny{ML}}} \equiv \frac{\pi \hbar\left(1-\operatorname{Re} \tilde{A}(T)+\frac{2}{\pi} \operatorname{Im} \tilde{A}(T)\right)}{2\left(\langle H\rangle-E_0\right)} \; .
\end{equation}

For evolution to a fully orthogonal state, say at an orthogonalisation time $T$, we would have $A(T)=0$. This implies that both the real and imaginary parts of $\tilde A(T)$ also vanish, giving in this special case
\begin{equation}
0\ge 1-\frac{2\bigl(\langle H\rangle-E_0\bigr)T}{\pi\hbar} \; ,
\end{equation}
leading to
\begin{equation}
T \ge \frac{\pi\hbar}{2\bigl(\langle H\rangle-E_0\bigr)} \;.
\end{equation}
One identifies from this the Margolus--Levitin characteristic time\,\cite{Margolus:1997ih} for full orthogonalisation of states as 
\begin{equation}
T^{\perp}_{\text{\tiny{ML}}} \equiv \frac{\pi\hbar}{2\bigl(\langle H\rangle-E_0\bigr)} \; .
\end{equation}

\printbibliography
\end{document}